%% file: main.tex
\documentclass[10pt, letterpaper, journal]{IEEEtran}
\usepackage{amsmath,amssymb,amsfonts}
\usepackage{algorithmic}
\usepackage[linesnumbered,ruled]{algorithm2e}
\usepackage{graphicx}
\usepackage{textcomp}
\usepackage{pifont}
\usepackage{xcolor}
\usepackage{threeparttable}
\usepackage{tabularx}
\usepackage{booktabs}
\usepackage{multirow}
\usepackage{colortbl}
\usepackage{enumitem}
\usepackage{subfigure}
\usepackage{amsthm}
\usepackage{pifont}
\usepackage{comment}
\usepackage{array} 

\usepackage{bigstrut,multirow,rotating}
\usepackage[numbers,sort&compress]{natbib}
\usepackage{tikz}
\usepackage[colorlinks, linkcolor=magenta, anchorcolor=green, citecolor=blue]{hyperref}
\usepackage[normalem]{ulem}

\usepackage{hhline}

\def\BibTeX{{\rm B\kern-.05em{\sc i\kern-.025em b}\kern-.08em
    T\kern-.1667em\lower.7ex\hbox{E}\kern-.125emX}}
    
\newcommand*{\circled}[1]{\lower.7ex\hbox{\tikz\draw (0pt, 0pt)%
    circle (.5em) node {\makebox[1em][c]{\small #1}};}}

\usepackage{amsmath, amssymb, amsthm}
\usepackage{mdframed}
\definecolor{shadecolor}{rgb}{0.94, 0.94, 0.94}

\newcounter{shadeddefinition}[section]
\renewcommand{\theshadeddefinition}{\thesection.\arabic{shadeddefinition}}
\newenvironment{shadeddefinition}[1][]
{\refstepcounter{shadeddefinition}
\begin{mdframed}[backgroundcolor=shadecolor, linewidth=0pt, innerleftmargin=6pt, innerrightmargin=6pt, innertopmargin=6pt, innerbottommargin=6pt]
\noindent\textbf{Definition \theshadeddefinition\ (#1).} }
{\end{mdframed}}

\usepackage{framed}  % For the framed environment
\newenvironment{finding}[1]
{\begin{framed}\noindent\textbf{#1} \itshape}
{\end{framed}}

\begin{document}

\title{\huge CIMinus: Empowering Sparse DNN Workloads Modeling and Exploration on SRAM-based CIM Architectures}

\author{Yingjie~Qi,
        Jianlei~Yang,~\IEEEmembership{Senior Member,~IEEE,}
        Rubing~Yang,
        Cenlin~Duan,
        Xiaolin~He,
        Ziyan~He,
        Weitao~Pan,~\IEEEmembership{Member,~IEEE,}
        and~Weisheng~Zhao,~\IEEEmembership{Fellow,~IEEE}
\thanks{
Manuscript received on October 2, 2024, revised on July 29, 2025, accepted on October 29, 2025.
This work was supported in part by the National Natural Science Foundation of China (Grant No. 62572036), the Beijing Natural Science Foundation (Grant No. L243031), the National Key R\&D Program of China (Grant No. 2023YFB4503704 and 2024YFB4505601), and the Academic Excellence Foundation of BUAA for PhD Students. \textit{The corresponding author is Jianlei Yang.}
}
\thanks{Y. Qi, J. Yang, R. Yang, X. He, are with School of Computer Science and Engineering, Beihang University, Beijing, 100191, China. E-mail: \url{jianlei@buaa.edu.cn}}
\thanks{C. Duan and W. Zhao are with School of Integrated Circuit Science and Engineering, Beihang University, Beijing, 100191, China.}
\thanks{Z. He and W. Pan are with School of Telecommunications Engineering, Xidian University, Xi'an, 710071, China.}
}

\maketitle

\input{texfiles/0-Abstract.tex}

\begin{IEEEkeywords}
Compute-In-Memory, Sparsity, Deep Learning, SRAM-CIM
\end{IEEEkeywords}

\IEEEpeerreviewmaketitle

\bstctlcite{IEEEexample:BSTcontrol}

\input{texfiles/1-Introduction.tex}
\input{texfiles/2-Motivation.tex}
\input{texfiles/3-Abstraction.tex}
\input{texfiles/4-Framework.tex}
\input{texfiles/5-Methodology.tex}

\input{texfiles/6-Evaluation}
\input{texfiles/7-Exploration.tex}
\input{texfiles/8-RelatedWorks}
\input{texfiles/9-Conclusion.tex}

{
\small
\bibliographystyle{IEEEtran}
\bibliography{ref}
}

\input{texfiles/bio.tex}

\end{document}

%% file: texfiles/0-Abstract.tex
\begin{abstract}

Compute-in-memory (CIM) has emerged as a pivotal direction for accelerating workloads in the field of machine learning, such as Deep Neural Networks (DNNs).
However, the effective exploitation of sparsity in CIM systems presents numerous challenges, due to the inherent limitations in their rigid array structures.
Designing sparse DNN dataflows and developing efficient mapping strategies also become more complex when accounting for diverse sparsity patterns and the flexibility of a multi-macro CIM structure.
Despite these complexities, there is still an absence of a unified systematic view and modeling approach for diverse sparse DNN workloads in CIM systems.
In this paper, we propose CIMinus, a framework dedicated to cost modeling for sparse DNN workloads on CIM architectures. 
It provides an in-depth energy consumption analysis at the level of individual components and an assessment of the overall workload latency.
We validate CIMinus against contemporary CIM architectures and demonstrate its applicability in two use-cases.
These cases provide valuable insights into both the impact of sparsity patterns and the effectiveness of mapping strategies, bridging the gap between theoretical design and practical implementation.

\end{abstract}

%% file: texfiles/1-Introduction.tex
\section{Introduction}\label{sec:introduction}

\IEEEPARstart{D}{ue} to the growth of computing power and the availability of big data, the field of machine learning has witnessed significant advancements in recent years.
The rapid development of intelligent applications has motivated many dedicated accelerator designs for Deep Neural Networks (DNNs)~\cite{chen2016eyeriss, jouppi2017datacenter}.
However, the ever-increasing scale and complexity of DNN models make conventional NN accelerators suffer from severe data movement overhead~\cite{gholami2024ai}.
To address this bottleneck, compute-in-memory (CIM) architectures have emerged as a promising solution, reducing data transfer overhead by integrating computation logic within memory technologies such as SRAM~\cite{kang2018multi, si202015, chih202116}, ReRAM~\cite{chi2016prime, shafiee2016isaac}, and MRAM~\cite{patil2019mram}.
Among these viable candidates, SRAM is widely adopted across industry and academia due to its faster write speeds, lower write energy consumption, and compatibility with existing logic technologies~\cite{jhang2021challenges}.

Alongside the evolution of CIM architecture designs, there is an increasing emphasis on software-hardware co-design approaches that integrate hardware capabilities with model-level optimizations.
Among co-design approaches, sparsity exploitation is one of the most popular strategies for improving energy efficiency~\cite{yang2019sparse, chu2020pim, yang2021auto, zheng2022flexible, giannoula2022sparsep}.
Although sparsity support has been widely adopted in conventional accelerators, the methods employed are not readily adaptable to CIM architectures.
Due to the constraints of rigid crossbar structures, the irregularities brought by sparse data structures often lead to under-utilization problems in CIM macros.
Designers also face a wide range of design choices when developing in-memory architectures for sparse DNN workloads.
For instance, while the multi-macro organization in contemporary CIM accelerators offers accommodation for more complex workloads~\cite{wang2022spcim}, it may also complicate the design of efficient sparse dataflows and their corresponding mapping strategies.

Prior studies have proposed several SRAM-based CIM designs to leverage DNN sparsity~\cite{yue202014, yue202115, sie2021mars, tu2022sdp}.
However, the domain of sparsity exploitation within CIM has not been thoroughly investigated, as the strategies for selecting sparsity patterns and mapping methods in these accelerators largely rely on heuristic or empirical approaches.
Such reliance may overlook optimization opportunities in both CIM architectures and DNN sparsity.
Particularly, finer-grained sparsity patterns in DNN weights tend to preserve model accuracy but pose significant challenges for efficient utilization.
Moreover, when accounting for the diversity of sparsity patterns available within the constraints of a multi-macro CIM framework, the design of sparse DNN dataflows and the formulation of mapping strategies also become exceedingly complex.
The absence of a systematic approach, combined with the complexities in evaluating sparse workloads on CIM architectures, poses significant challenges for exploring the space of algorithm-hardware co-design.

This paper introduces CIMinus, a cost modeling framework that enables designers to efficiently navigate the expansive co-design space.
CIMinus provides an efficient interface for both software and hardware descriptions, accurately estimating the energy consumption and latency of sparse workloads on CIM architectures while simplifying the user effort involved.
To facilitate the exploration of various sparsity patterns and their impact on CIM performance, CIMinus also incorporates a pruning workflow that generates sparse DNN weights in accordance with our proposed \textit{FlexBlock} sparsity abstraction.
Compared with the reported results in recent sparse CIM accelerators, CIMinus can accurately estimate the speedups and energy savings brought by DNN sparsity within the error margin of 5.27\%.
We also demonstrate the capabilities of CIMinus through two use-cases exploring sparsity exploitation and mapping strategies.
The contributions of this paper include:
\begin{itemize}
    \item We present \textit{FlexBlock} sparsity, an expressive abstraction for describing sparsity patterns under structural constraints, streamlining the process of integrating sparse DNNs with CIM architectures.
    \item We propose CIMinus, an expedient cost modeling framework for sparse DNN workloads on SRAM-based digital CIM architectures and validate it against contemporary CIM architectures.
    \item We demonstrate the practical applicability of CIMinus through two illustrative use-cases, bridging theoretical design and practical implementation in DNN sparsity and mapping exploration.
\end{itemize}

The rest of this paper is organized as follows. 
Sec.~\ref{sec:motivation} provides background on SRAM-based CIM architectures and their challenges in sparsity exploitation.
Sec.~\ref{sec:sparsity} details the constraints and hardware support for sparsity in CIM architectures and introduces the \textit{FlexBlock} sparsity abstraction.
Sec.~\ref{sec:framework} and Sec.\ref{sec:methodology} present the CIMinus framework and its modeling methodology, followed by their validation and runtime analysis in Sec.~\ref{sec:evaluation}. 
Sec.~\ref{sec:exploration} explores the impact of sparsity patterns and mapping strategies, Sec.~\ref{sec:related} discusses the related works, and concluding remarks are given in Sec.~\ref{sec:conclusion}.

%% file: texfiles/2-Motivation.tex
\section{Background and Motivation}\label{sec:motivation}

In this section, we first discuss the major design trends of SRAM-based CIM to establish the groundwork of our study.
We explain the shift from analog to digital in the computing paradigm, and then highlight the challenges associated with sparsity exploitation in CIM, which motivate the development of our proposed framework.

\subsection{Design Trends in SRAM-based CIM}

\begin{figure}[t]
    \centering
    \subfigure[Architectural overview of analog and digital CIM.]{
        \centering
        \includegraphics[width=0.48\textwidth]{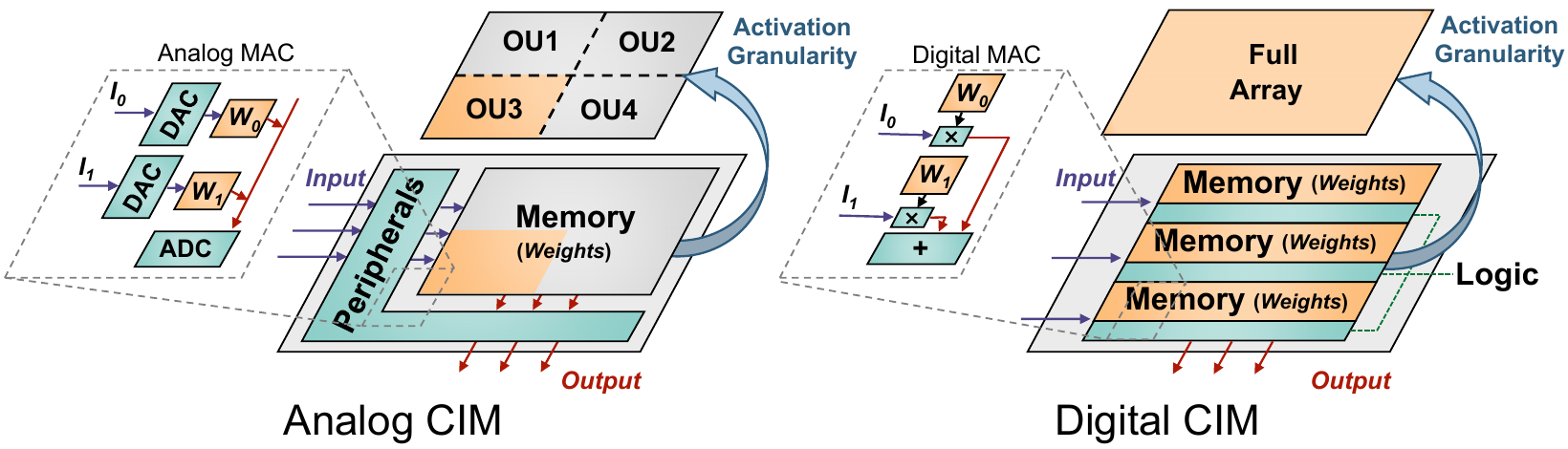}
        \label{fig:motivation:CIM}
    }
    \subfigure[Sparsity patterns utilized by recent CIM designs in literature.]{
        \centering
        \includegraphics[width=0.47\textwidth]{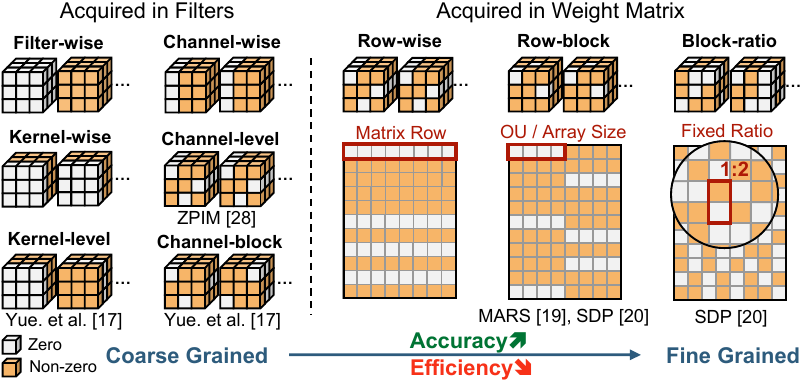}
        \label{fig:motivation:sparsity}
    }
    \caption{Comparative overview of CIM architectures and sparsity patterns.}
    \label{fig:motivation}
\end{figure}

CIM architectures perform Multiply-Accumulate (MAC) operations in close proximity to or directly within memory arrays, reducing the data movement bottleneck of traditional von Neumann architectures.
Many of the initial CIM designs in literature (such as~\cite{si202015, su202015, su202116}) implement analog MAC operations using current or voltage signals.
The inherent analog deviations and the overhead associated with analog-to-digital converters (ADCs) significantly constrain the accuracy and utilization of the memory array.
This limitation results in a confined activation granularity within the array, often termed as an Operation Unit (OU).
Consequently, a clear trend in CIM architecture design is the shift towards digital implementations.
Digital CIM integrates digital MAC logic into SRAM cells, effectively avoiding the challenges posed by analog non-idealities~\cite{kim20191, chih202116, yan20221, guo202328nm, duan2024ddc, duan2024towards}.
As depicted in Fig.~\ref{fig:motivation:CIM}, digital CIM implementations offer the potential to activate all rows simultaneously, significantly enhancing computational parallelism and array utilization.

To support a wider range of intelligent applications, another notable trend in CIM design involves integrating more macros into the CIM systems.
This approach not only increases the capacity of the in-memory architecture, but also improves the overall system performance and efficiency~\cite{su202116, wang2022spcim}.
Moreover, as the complexity of DNN workloads keeps increasing, sparsity exploitation has also become an important strategy for improving the computational efficiency in CIM systems.
By leveraging sparsity introduced through pruning techniques, CIM systems can significantly reduce both storage and computation requirements, enabling them to efficiently handle larger and more complex DNN models.
While efforts have been made to utilize sparsity in analog SRAM-based CIM architectures~\cite{yue202014, yue202115, kim2021z, wang2022spcim} as well as in other technology-based CIMs~\cite{yang2019sparse, chu2020pim, yang2021auto, zheng2022flexible, giannoula2022sparsep}, the full array activation constraint in digital CIM architectures greatly limits the efficient utilization of sparse workloads.
This has led to relatively few studies~\cite{sie2021mars, tu2022sdp} focusing on sparsity exploration in digital CIM.

\subsection{Challenges and Design Space}\label{sec:motivation:designspace}

Performing sparse computations in-memory is not without challenges.
To fully utilize the high parallelism and energy efficiency promised by CIM designs, the executed workloads need to exhibit a certain degree of regularity or structure.
This requirement is particularly critical in digital CIM architectures, where the activation granularity spans the entire array.
Consequently, sparsity patterns in DNN workloads must adhere to structural constraints for effective execution in these systems.

As illustrated in Fig.~\ref{fig:motivation:sparsity}, the sparsity patterns employed by recent CIM architectures involve a wide range of granularities, which can be categorized as patterns directly acquired from pruning filters, and those derived from pruning within reshaped weight matrices.
Coarser-grained sparsity patterns are typically attained by pruning values from specific dimensions of DNN weights, while finer-grained patterns are obtained in a block-wise fashion.
The blocks are typically configured to either have matching sizes to the OU/array in CIM architectures~\cite{sie2021mars, wang2022spcim}, or to maintain a fixed ratio of non-zero values~\cite{tu2022sdp}.
Although finer granularity in sparsity patterns can preserve model accuracy, it also inevitably introduces additional storage and indexing overhead.
Furthermore, while most CIM systems adopt a weight stationary mapping approach, where DNN weights are mapped onto CIM macros for data reuse, the introduction of sparsity can lead to workload imbalances among these macros.
The random nature of sparsity often prevents a perfect mapping of compressed weights onto multiple CIM macros, resulting in reduced spatial parallelism and macro under-utilization.

To fully leverage the flexibility of multi-macro digital CIM systems and the advantages of sparsity, it is essential to explore a broad design space of software/hardware co-design.
However, many existing CIM accelerator designs are often tailored to specific patterns of DNN sparsity.
Moreover, modeling tools for CIM designs~\cite{peng2019dnn+, zhu2023mnsim, sun2023analog, andrulis2024cimloop} emphasize on accurately emulating the hardware designs, primarily targeting dense workloads.
As a result, there still lacks a systematic view on the execution of diverse sparse DNN workloads in CIM systems, leaving several key questions unaddressed, some of which we list below.
The aim of our CIMinus framework is to provide a tool that facilitates easy navigation through this complex design space, filling the gap in the existing body of research.

\begin{itemize}
    \item What types of sparse workload are best suited for CIM?
    \item How to efficiently map a sparse workload to a CIM architecture to avoid under-utilization?
\end{itemize}

%% file: texfiles/3-Abstraction.tex
\section{Sparsity in CIM Architectures}\label{sec:sparsity}

Sparsity exploitation is a crucial technique for improving efficiency of DNN acceleration on CIM architectures. 
However, the structural characteristics of CIM arrays impose unique challenges and constraints on sparsity exploitation. 
In this section, we explore these challenges in detail,  examining both the structural limitations of CIM arrays and the hardware support required for sparsity in model inputs and weights.
We then introduce the \textit{FlexBlock} sparsity abstraction as a solution to effectively represent the diverse sparsity patterns in CIM-based DNN accelerators.

\subsection{Structural Limitations of CIM Arrays}\label{sec:sparsity:limitation}

\begin{figure}[t]
    \centering
    \includegraphics[width=\linewidth]{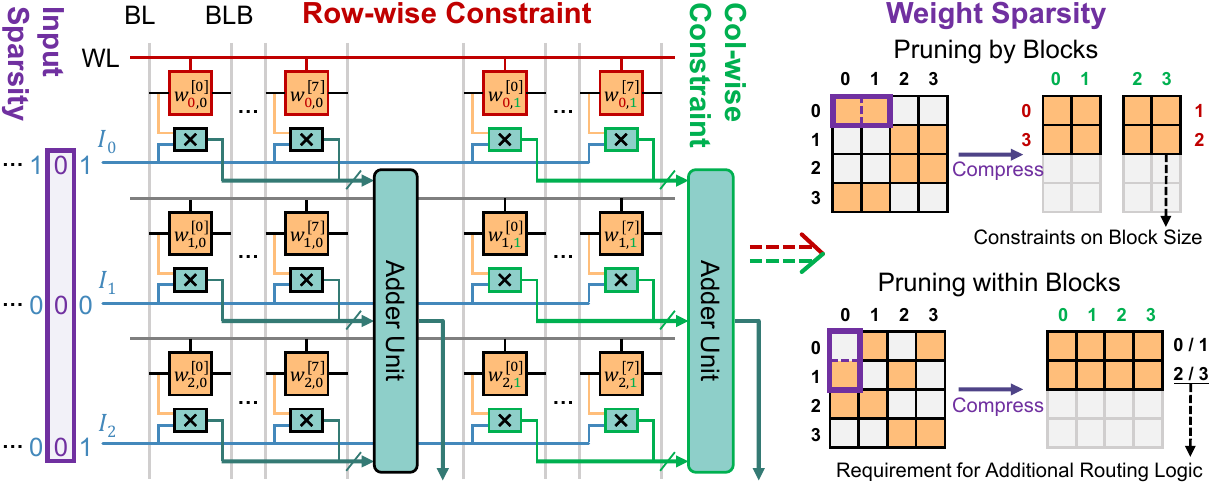}
    \caption{Structural constraints of CIM architectures on sparsity exploitation.}
    \label{fig:cim-constraints}
\end{figure}

In order to maximize in-memory computation benefits, DNN weights are typically reshaped from their original multi-dimensional structures into two-dimensional matrices and stored stationarily within the memory arrays.
The computations are executed in a bit-serial manner, where input values are decomposed into individual bits and fed sequentially to the array.
As shown in Fig.~\ref{fig:cim-constraints}, during computation, corresponding rows within each sub-array are activated simultaneously according to wordline signals, enabling row-wise parallelism.
The partial sums resulting from these row-wise operations are then accumulated along the bitline direction.
Although the crossbar structure in CIM architectures enables efficient parallel computing, it also imposes rigid requirements.
The row-level activation mechanism demands inputs to be broadcast across all elements in each row, constraining the row-wise data indices to be identical.
Similarly, the accumulation process requires the alignment of the output data indices across rows to ensure correct functionality.
Beyond these structural constraints, SRAM arrays typically can store only a fraction of DNN layer weights, often requiring designers to employ multi-macro architectures or develop complex dataflow mappings.

\subsection{Sparsity Support in CIM}\label{sec:sparsity:support}
To reduce both computation and storage requirements in DNN acceleration, many modern CIM architectures leverage the sparsity present in neural network computations.
This includes utilizing input sparsity by skipping the zero bits in activations during bit-serial processing, and exploiting weight sparsity by compressing weight matrices after model pruning.
However, the irregular nature of sparsity often leads to a mismatch between the flexible zero distributions and the rigid crossbar structure, resulting in under-utilization of the available parallelism and computational resources. 
For input sparsity, computations can only be skipped when identical bit positions are zero across all inputs processed by the activated rows in a CIM array.
For weight sparsity, common patterns such as unstructured or fine-grained sparsity inherently violate the row-wise and column-wise constraints imposed by CIM architectures, which largely negates the computational and storage benefits that sparsity provides to address the SRAM array capacity constraint.

These challenges require a co-design approach where CIM architectures include dedicated logic for sparsity support, and sparsity patterns are structured to match the rigid hardware constraints.
To support input sparsity, CIM architectures must identify when bit positions are zero across inputs during the bit-serial conversion and dynamically skip these ineffectual computations.
To effectively utilize weight sparsity, the sparsity patterns must ensure that compression can preserve the alignment between matrix indices and the corresponding CIM array coordinates.
This requires the weights to be pruned in a block-wise fashion, which can be realized through two approaches: pruning entire blocks that conform to the hardware dimensions, or consistently pruning within each block to obtain a uniform compressed shape.
After model pruning and compression, these structured weights can be stored densely within CIM arrays, as long as the architecture maintains the offline-generated indexing information needed to route inputs to the appropriate array rows.

\subsection{FlexBlock Sparsity Abstraction}\label{sec:sparsity:flexblock}

While sparsity in CIM must adhere to structural constraints, practical implementations employ diverse sparsity patterns with a wide range of granularities.
To effectively represent structurally constrained sparsity patterns for two-dimensional weight matrices, we introduce \textit{FlexBlock}, a flexible and unified sparsity abstraction that serves as the foundation for the CIMinus modeling framework.
The \textit{FlexBlock} abstraction is designed to accommodate diverse sparsity patterns by representing them as a composition of multiple block-based sparsity patterns, which can be defined as follows.

\begin{shadeddefinition}[\textit{FlexBlock} Sparsity]
Given a matrix $W \in \mathbb{R}^{M \times N}$, \textit{FlexBlock} sparsity is defined as a set of block-based sparsity patterns $\mathcal{B} = \{B_1, B_2, \dots, B_k\}$, where $B_i$ represents the $i$-th sparsity pattern applied to $W$. 
Each $B_i$ can be either a \textit{FullBlock} or \textit{IntraBlock} sparsity pattern with block size $s_i = (m_{i}, n_{i})$, where $0 < m_i \leq M$, $0 < n_i \leq N$, $m_i \times n_i > 1$, and sparsity ratio $r_i \in (0, 1)$.
\end{shadeddefinition}

The \textit{FlexBlock} abstraction leverages two primary types of block-based sparsity patterns: \textit{FullBlock} sparsity and \textit{IntraBlock} sparsity. 
\textit{FullBlock} sparsity refers to the case where all elements within a block are pruned, resulting in a block of zeros. 
This type of sparsity pattern is commonly observed in coarser-grained pruning strategies, where entire blocks or submatrices are removed. 

\begin{shadeddefinition}[\textit{FullBlock} Sparsity]
\textit{FullBlock} sparsity $B_F$ is a type of sparsity pattern where all elements within designated blocks of size $m \times n$ in matrix $W$ are zero.
The sparsity ratio $r \in (0, 1)$ defines the proportion of zero blocks within the matrix, such that the number of non-zero blocks is given by $\Phi = \lfloor (1 - r) \cdot \frac{M}{m} \cdot \frac{N}{n} \rfloor$.

Let $I_B = \{(i_1, j_1), (i_2, j_2), \dots, (i_{\Phi}, j_{\Phi})\}$ be the set of indices of the non-zero blocks in $W$, where each $(i_l, j_l)$ represents the top-left corner of the $l$-th non-zero block.
For matrix $W$ to exhibit \textit{FullBlock} sparsity, each block $W_{i:i+m, j:j+n}$ in $W$ must satisfy the following condition:
\[
W_{i:i+m, j:j+n} =
\begin{cases}
\mathbf{0}_{m \times n}, & \text{if } (i, j) \notin I_B \\
W^O_{i:i+m, j:j+n}, & \text{if } (i, j) \in I_B
\end{cases},
\]
where $W^O$ denotes the original matrix before sparsification.
\end{shadeddefinition}

In contrast, \textit{IntraBlock} sparsity allows for more fine-grained pruning within each block, where elements are pruned according to a preset ratio and a series of predefined patterns. 
This enables the representation of more complex and irregular sparsity patterns within each block.

\begin{shadeddefinition}[\textit{IntraBlock} Sparsity]
\textit{IntraBlock} sparsity is a type of sparsity pattern where the arrangement of non-zero elements within each $m \times n$ block of matrix $W$ is determined by a sparsity ratio $r \in (0, 1)$ and a pattern set $\mathcal{P}$.
The sparsity ratio $r$ defines the proportion of zero elements within each block, such that the number of non-zero elements is given by $\phi = \lfloor (1 - r) \cdot m \cdot n \rfloor$.
The pattern set $\mathcal{P} = \{P_1, P_2, \dots, P_k\}$ is composed of binary masks, each of size $m \times n$, that specify the locations of the $\phi$ non-zero elements within a block.

For matrix $W$ to exhibit \textit{IntraBlock} sparsity, each block $W_{i:i+m, j:j+n}$ in $W$ must satisfy the following condition:
\[
\exists P \in \mathcal{P}, \text{ s.t. } W_{i:i+m, j:j+n} \odot P = W_{i:i+m, j:j+n} ,
\]
where $\odot$ denotes element-wise multiplication.
\end{shadeddefinition}

By composing \textit{FullBlock} and \textit{IntraBlock} sparsity patterns in a specified order, \textit{FlexBlock} provides an expressive abstraction for weight sparsity that enables the modeling and exploration of diverse sparsity configurations on CIM architectures.

\begin{figure}[t]
    \centering
    \includegraphics[width=\linewidth]{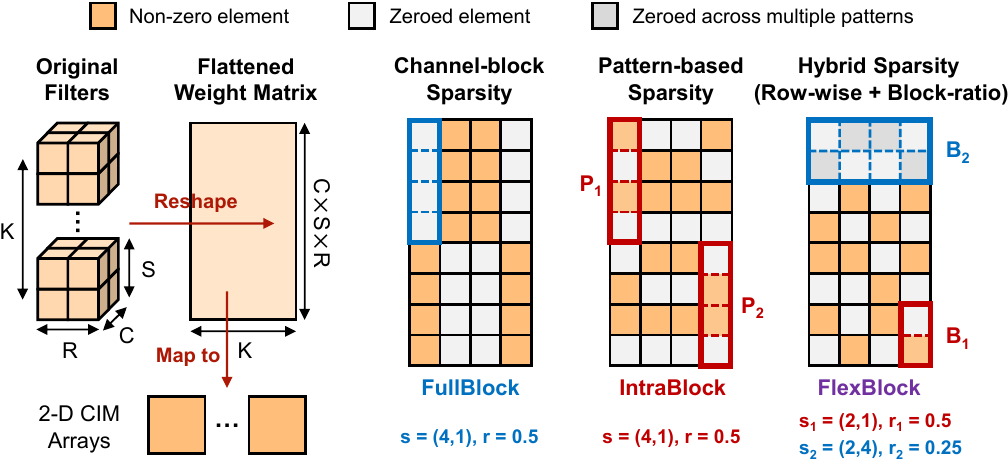}
    \caption{Examples of sparsity pattern representation in reshaped two-dimensional weight matrices using the \textit{FlexBlock} abstraction.}
    \label{fig:flexblock}
\end{figure}

\subsection{Sparsity Description with \textit{FlexBlock}}
As illustrated in Fig.~\ref{fig:flexblock}, the original three-dimensional weight filters must be reshaped into two-dimensional matrices before mapping onto CIM arrays.
\textit{FlexBlock} can represent the diverse sparsity patterns found in recent CIM designs operating on these reshaped matrices.
For instance, the channel-block sparsity pattern utilized in \cite{yue202014}, where weights along the channel direction are pruned according to a fixed block size, can be represented with \textit{FullBlock} sparsity on the weight matrix flattened in a channel-major order.
Similarly, the pattern-based sparsity employed in SegPrune~\cite{zheng2022flexible}, which follows a predefined set of block-wise sparsity patterns, can be captured using \textit{IntraBlock} sparsity with the corresponding pattern set.
\textit{FlexBlock} can also easily depict the hybrids and variants of commonly exploited sparsity patterns, such as the combination of block-ratio sparsity and row-wise sparsity in SDP~\cite{tu2022sdp}.

\begin{figure*}
    \centering
    \includegraphics[width=0.98\linewidth]{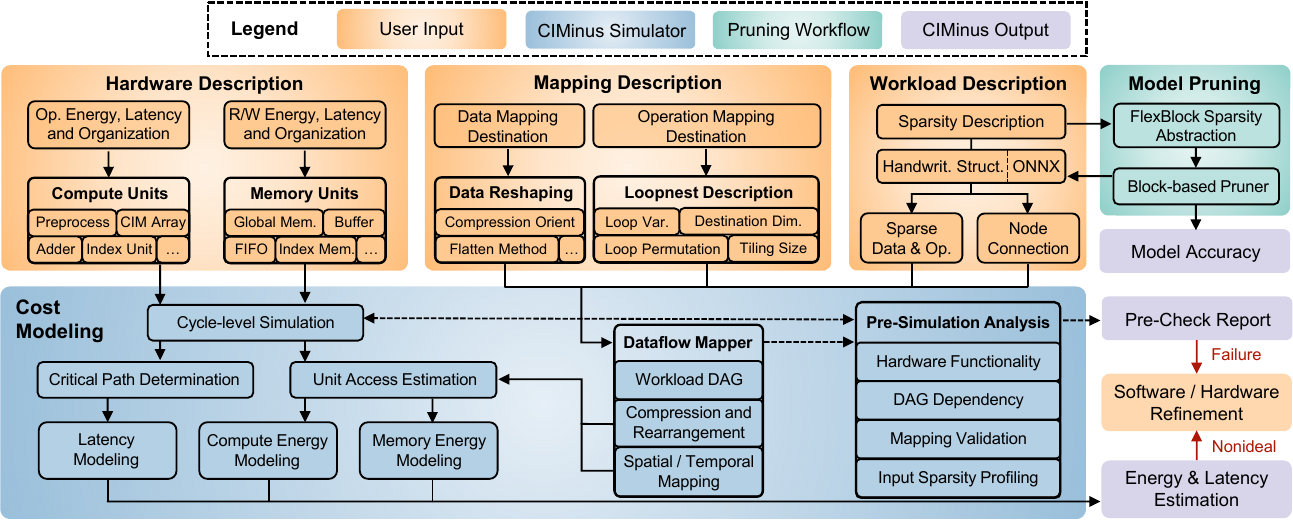}
    \caption{Overview of the CIMinus framework.}
    \label{fig:overview}
\end{figure*}

However, while the \textit{FlexBlock} abstraction can theoretically represent any given sparsity pattern in two-dimensional matrices, 
the patterns must satisfy several constraints to ensure efficient CIM mapping in practice.
The block sizes of \textit{FullBlock} patterns should be integral multiples of the hardware dimensions to avoid misalignment, while \textit{IntraBlock} patterns must be column-wise one-dimensional blocks to maintain uniform compressed shapes and ensure correct accumulation along the bitline direction.
The efficient accommodation of these patterns also requires CIM architectures to employ specialized hardware support.
For workloads with \textit{FullBlock} sparsity, the computations within each CIM array remain identical to dense implementations.
The accelerator only needs to store block-level indices that direct inputs to the corresponding array rows and coordinate the accumulation of partial sums. 
In contrast, \textit{IntraBlock} sparsity requires finer-grained support.
Since column-wise compression maps weight elements from multiple rows into a single array row, the accelerator must broadcast multiple inputs to each row and maintain element-level indices to select the correct input for each weight. 

Furthermore, the composition of sparsity patterns also needs to adhere to certain constraints.
To prevent excessive fragmentation in the final matrix, the block size of the coarser-grained pattern must be an integral multiple of its finer-grained counterpart.
In addition, we choose to limit pattern composition to a maximum of two in CIMinus based on the following considerations.
On the one hand, with adjustable block sizes and sparsity ratios, the two types of sparsity patterns and their combination can already capture the mainstream sparsity patterns in existing CIM designs. 
On the other hand, adding patterns beyond two can only provide diminishing returns.
Stacking multiple \textit{IntraBlock} patterns would significantly expand the range of rows that are mapped to an array row, increasing the routing complexity and indexing overhead.
Meanwhile, composing multiple \textit{FullBlock} patterns can only produce pruning results that are the mathematical subset of its finest-grained pattern, due to the integral multiple constraint.
Therefore, the two-pattern limit represents a practical balance between expressiveness and complexity.
These practical constraints ensure \textit{FlexBlock} patterns can be efficiently mapped to CIM arrays while maintaining the flexibility to represent both existing and emerging sparsity configurations.

%% file: texfiles/4-Framework.tex
\section{CIMinus Framework}\label{sec:framework}

In this section, we first discuss the design rationale behind CIMinus, highlighting its key objectives and intended use cases. 
We then provide an overview of the framework's main components and their interactions. 
Finally, we detail the programming interface of CIMinus, showcasing its flexibility and ease of use through illustrative examples.

\subsection{Design Rationale of CIMinus}
CIMinus is designed to enable informed decision-making for sparsity exploitation in the early stages of CIM development, when the design space is most flexible but the cost of exploring different configurations through circuit-level simulation or manual analysis is prohibitive.
Compared to existing dense-focused evaluation tools for CIM~\cite{peng2019dnn+, zhu2023mnsim, sun2023analog, andrulis2024cimloop}, CIMinus provides a unified environment spanning from DNN model pruning to system-level cost modeling, where changes to sparsity patterns, mapping strategies, or hardware configurations can be evaluated as a whole.
This tool is intended for use once preliminary component designs have been developed, as CIMinus requires their performance characteristics to provide accurate system-level estimations.
For instance, CIMinus can help designers to evaluate the efficacy of utilizing their CIM architecture for a given sparse workload, or identify the most effective mapping strategy to optimize system performance and reduce overall overhead.
However, it is important to understand that CIMinus is \textit{not} a synthesis tool for generating energy profiles of individual accelerator components, nor is it a specific CIM accelerator implementation.
Instead, CIMinus is designed to complement tools like High-Level Synthesis (HLS), which can be used to create accelerator designs for system-level evaluation and exploration.

\subsection{Framework Overview}\label{sec:framework:overview}

The fundamental approach for evaluating energy consumption is to aggregate the product of the access count of each hardware unit with its respective per-access energy consumption.
Similarly, determining the critical path among hardware components is essential for overall latency estimation.
Therefore, the primary objective of CIMinus is to develop a modeling methodology for these metrics, complemented by a pruning workflow that leverages the \textit{FlexBlock} abstraction and a user-friendly programming interface tailored to minimize user effort.
A comprehensive illustration of the CIMinus framework is shown in Fig.~\ref{fig:overview}.
Here we provide a brief overview of the programming interface, pruning workflow, and modeling methodology.
Detailed discussions of these components can be found in Sec.~\ref{sec:framework:interface}, Sec.~\ref{sec:framework:pruning}, and Sec.~\ref{sec:methodology}, respectively.

\textbf{Interface.} In CIMinus, the hardware components of CIM architectures are abstracted as a collection of basic compute and memory units, while the DNN workloads are depicted through a directed acyclic graph (DAG).
The sparsity patterns within these workloads are encapsulated with a description corresponding to the proposed \textit{FlexBlock} abstraction.
Moreover, the mapping method can be described by a versatile mapping template, allowing for both the spatial and temporal mapping of loopnest-based dataflows.
By decoupling the description of DNN workloads, CIM architectures, and the corresponding mapping strategy, this interface substantially streamlines the iterative process of system exploration.

\textbf{Pruning Workflow.} 
CIMinus incorporates a pruning workflow that leverages the \textit{FlexBlock} abstraction to generate diverse sparsity patterns tailored for CIM architectures.
This workflow allows users to define and apply pruning strategies at both coarse- and fine-grained levels.
The pruning process involves calculating the importance of each block or each element within a block based on user-specified criteria, and generating pruning masks to zero out the selected elements.
Seamlessly integrated with the CIMinus framework, the pruning workflow enables users to explore and evaluate the impact of sparsity patterns on the performance and energy efficiency of CIM systems.

\textbf{Modeling Methodology.} To accurately model the energy consumption of CIM systems for a specific workload, CIMinus requires users to provide the per-access energy or per-cycle energy for their compute and memory units.
Such energy statistics are generally obtainable through ASIC synthesis flows or with the aid of memory modeling tools like PCACTI~\cite{shafaei2014fincacti}.
Prior to simulation, CIMinus conducts pre-simulation analysis that includes functional verification and input sparsity profiling.
First, the verification ensures the validity and consistency of hardware, workload, and mapping specifications. 
Additionally, for configurations with input sparsity, CIMinus performs model inference on dataset samples to extract the activations and estimate the ratio of skippable computations, since the sparsity in activations is processed dynamically during execution and varies with each input. 
Based on these analyses, CIMinus either reports verification failures or proceeds to deliver comprehensive results including overall system latency and detailed energy breakdown.

\subsection{Programming Interface}\label{sec:framework:interface}

The programming interface of CIMinus includes the declarative descriptions for three domains: DNN workload, hardware unit, and mapping method.

\begin{figure}[t]
    \centering
    \subfigure[Example of Sparse DNN workload description.]{
        \centering
        \includegraphics[width=0.99\linewidth]{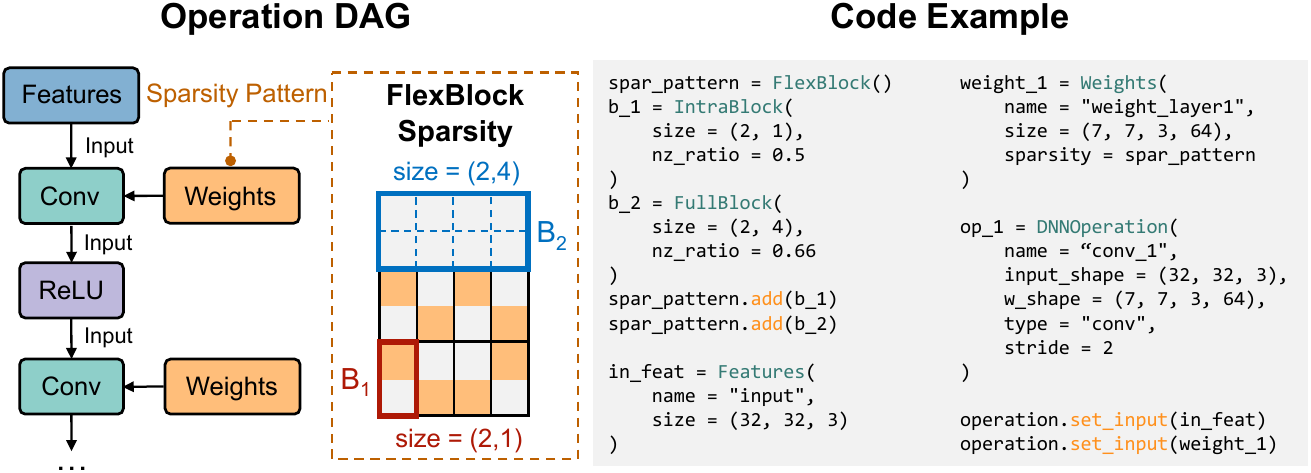}
        \label{fig:interface:interface_sparsity}
    }
    \subfigure[Example of hardware description.]{
        \centering
        \includegraphics[width=0.99\linewidth]{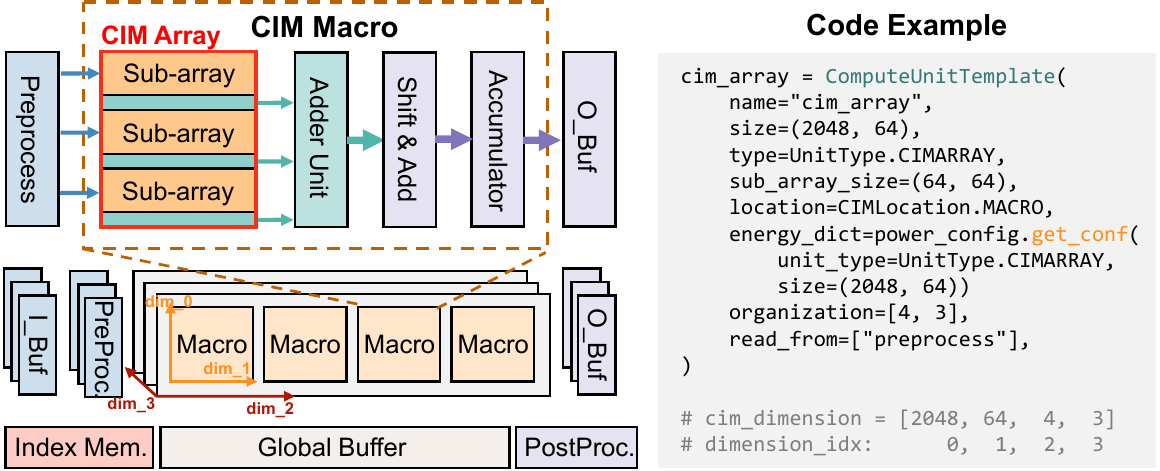}
        \label{fig:interface:CIM}
    }
    \subfigure[Example of mapping description.]{
        \centering
        \includegraphics[width=0.99\linewidth]{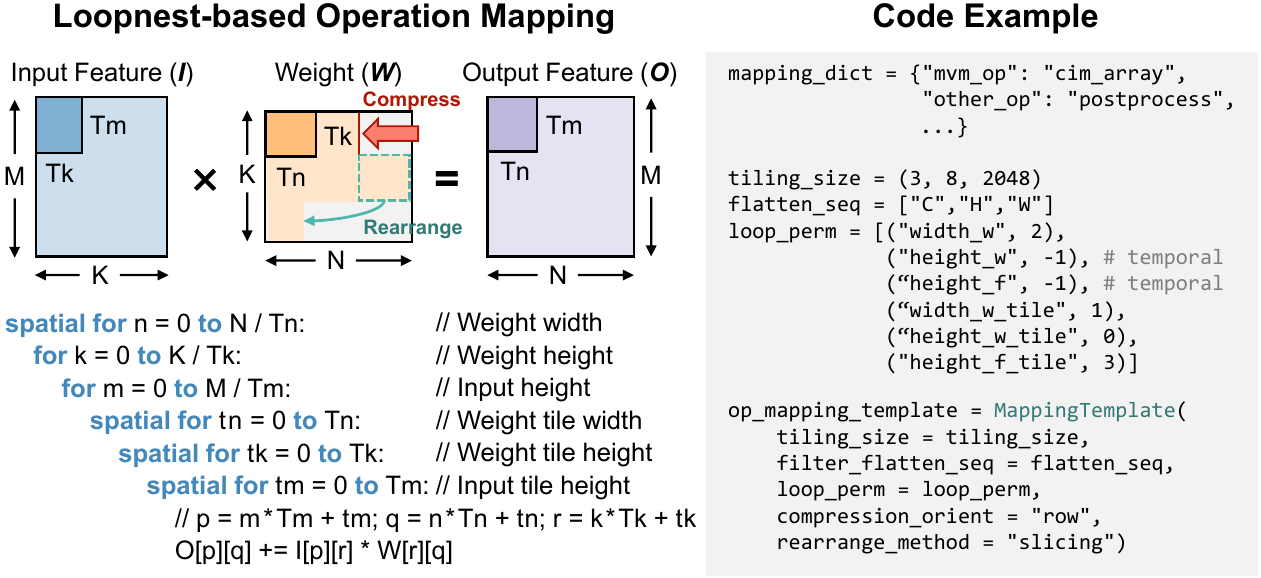}
        \label{fig:interface:mapping}
    }
    \caption{Describing sparse DNN workloads on an example CIM architecture design with the programming interface in CIMinus.}
    \label{fig:interface_hardware_mapping}
\end{figure}

\textbf{Workload Description.}
As illustrated in Fig.~\ref{fig:interface:interface_sparsity}, CIMinus represents a sparse DNN workload as a DAG, where the nodes represent data and operations, and the edges indicate the input/output relationship among them.
For DAG construction, CIMinus offers the flexibility of directly importing information from DNN models stored in the ONNX format.
The import process automatically extracts the dimensions of each operation and its corresponding weight and feature data, connecting them based on their dependencies.
Alternatively, users can also opt to utilize the workload description interface, manually specifying the dimension information of each node and linking each node to its respective input nodes.
This flexibility allows users to tailor the workload description to their specific evaluation needs, from preliminary exploration to comprehensive analysis.

CIMinus additionally provides users with a comprehensive set of description interfaces aligned with our proposed \textit{FlexBlock} sparsity abstraction.
Users are required to input the block size, sparsity ratio, and specify the pattern set for \textit{IntraBlock} patterns if necessary.
When the pattern set is not specified, it will default to all available patterns based on the provided block size and sparsity ratio.
This sparsity description is then forwarded to the pruning workflow to obtain the sparse data masks used in CIMinus simulation.
For user-defined DNN workloads, CIMinus will auto-generate a randomized sparsity mask in accordance with the provided pattern description.

\textbf{Hardware Description.}
In CIMinus, hardware descriptions are categorized into compute units and memory units. 
Fig.~\ref{fig:interface:CIM} presents a high-level example of a CIM accelerator architecture, along with a sample description of CIM arrays within this architecture.
A CIM architecture typically comprises a series of CIM macros, global and local buffers, alongside other units responsible for pre-/post-processing and sparsity support.

\ding{172} \underline{Compute Units.}
The core of any CIM hardware lies in the design of CIM macros, which are composed of several types of compute units. 
As shown in Fig.~\ref{fig:interface:CIM}, the CIM array in each macro consists of multiple sub-arrays that perform bit-serial MAC operations. 
These arrays are supported by additional compute units, such as adder units, shift-adders, and accumulators, which collectively enable the efficient execution of matrix-vector multiplications (MVM). 
In addition, pre-processing units convert input features into a bit-serial format suitable for CIM macros, while post-processing units handle computations such as activation, pooling, and residual connections.
These units work in tandem with the CIM macros to efficiently execute the complete sparse DNN workload.

\ding{173} \underline{Memory Units.}
The CIM accelerator architecture comprises several types of memory units, including global buffers, local buffers and index memories.
Specifically, global buffers are primarily used to store weights and input/output feature data for the CIM macros. 
On the other hand, local buffers directly connected to CIM macros provide intermediate storage during computations.
CIMinus also offers flexibility in modeling different memory configurations and designs.
Depending on the design specifications, weights and features can be stored either together in a single global buffer or in separate dedicated buffers.
Additionally, CIMinus supports the modeling of advanced memory structures, such as ping-pong buffers, to accommodate diverse design requirements. 

\ding{174} \underline{Sparsity Support Units.}
In CIMinus, the hardware units for sparsity support are described with the existing compute and memory unit interfaces.
For \textit{FlexBlock}-based weight sparsity, index memories are required to store the block and element indices, with capacity requirements automatically determined based on the sparsity configuration.
When pruning with \textit{FullBlock} patterns causes column-wise misalignment, additional accumulator units are required to handle the irregular partial sum aggregation. 
For \textit{IntraBlock} patterns, multiplexer-based indexing units are placed between preprocessing units and CIM macros, responsible for selecting the appropriate input for each weight element from multiple broadcasted inputs.
When input sparsity is enabled, the pre-processing units include non-zero bit position detection logic based on OR-gates that generates the bit-mask for each bit position, and leading one detection logic for skipping zero computations.

CIMinus provides a comprehensive parameterized description of the digital CIM paradigm, enabling users to specify the key characteristics of various hardware units. 
From the user perspective, describing a hardware unit involves specifying its energy information, dimension, the connection between units, as well as the organization of CIM macros and the associated units.
Users are required to input both the dimension and energy information for each unit, including dynamic energy per access and static energy per cycle.
CIMinus automatically infers the number of units required based on the CIM array size, unit size, and the organization parameter, simplifying the hardware description process. 
In CIMinus, the \texttt{organization} parameter plays a crucial role in describing the layout of hardware units. 
By providing a variable-length list that specifies the dimensions of the unit organization, CIMinus enables users to customize the hardware layout according to specific design requirements.
This flexibility enables CIMinus to adapt to different architectural styles and provide support for reconfigurable designs. 
Furthermore, CIMinus also allows users to specify the location of each unit, whether inside or outside the CIM macro, in order to ensure an accurate depiction of the overall architecture. 

\textbf{Mapping Description.}
The mapping of sparse DNN workloads onto CIM architectures involves reshaping the sparse weight data and determining the mapping destination of each operation.
Fig.~\ref{fig:interface:mapping} showcases an example of the mapping description for the weights of an operation to the CIM arrays illustrated in Fig.~\ref{fig:interface:CIM}.

\ding{172} \underline{Data Reshaping.}
CIMinus provides a comprehensive mapping description that outlines the process of compressing and rearranging the sparse weights before mapping them onto the CIM arrays.
The three-dimensional weight filters are first flattened into two-dimensional matrices according to the flattening sequence, which specifies the order in which dimensions are collapsed.
Users can then define the compression orientation as either row-wise or column-wise, tailoring the reshaping process to the specific sparsity pattern and CIM array structure.
After compression, CIMinus adjusts the compressed dimensions to align with the tile size, adding padding when necessary to ensure a seamless fit.
The tile size, assigned by user, represents the granularity of data partitioning and processing in the CIM architecture.
In cases where compression leads to ragged shapes, CIMinus also offers rearrangement methods to enhance the spatial utilization of CIM macros.
These methods include equalizing the compressed matrix dimensions through padding or slicing.
CIMinus supports rearranging the compressed matrix according to the user-defined slice size along the row or column direction.

\ding{173} \underline{Operation Mapping.}
For mapping the operations in the user-defined workload, users are required to specify the destination of each type of operation and data in \texttt{mapping\_dict}.
Operations involving MVM, such as convolutional (Conv) and fully connected (FC) layers, are mapped onto the CIM macros, while other operations are mapped onto the post-processing unit.
For MVM-based operations, CIMinus employs a multi-level loopnest format to effectively represent the computation process.
Each loop in the loopnest corresponds to a specific dimension in the reshaped weight matrix, input feature matrix, or tiled sub-matrices, establishing a direct correlation between the loop structure and the data organization.

CIMinus supports both temporal and spatial mapping of loops, where temporally mapped loops are executed sequentially, and spatially mapped loops are executed in parallel.
For spatial mapping, each loop is assigned to a particular dimension within the CIM arrays or their organizational layout, enabling efficient utilization of the parallel computing resources.
The mapping process either unrolls the weight matrix for weight loops, loading different weights into the CIM arrays at each iteration, or duplicates it for feature loops, allowing data reuse and reducing memory accesses.
The flexibility in mapping description allows for fine-grained control over the distribution of computations and data movement across the CIM architecture.

\subsection{Pruning Workflow}\label{sec:framework:pruning}

To explore the diverse sparsity patterns in CIM architectures, CIMinus incorporates a pruning workflow based on the \textit{FlexBlock} sparsity abstraction introduced in Sec.~\ref{sec:sparsity}.
This workflow enables users to define and apply block-based pruning strategies at both coarse-grained and fine-grained levels, aligning with the \textit{FullBlock} and \textit{IntraBlock} sparsity types in \textit{FlexBlock} sparsity abstraction.

The pruning workflow consists of two primary components: block representation and pruning strategies.
The block representation component allows users to specify the properties of each sparse block, such as shape, sparsity ratio, and block type, along with an optional pattern set for \textit{IntraBlock} sparsity.
It then generates a corresponding sparse mask based on the provided configuration.
The pruning strategies component determines which blocks or elements to prune based on calculated loss values using a specified pruning criterion, such as the magnitude of weights (L1 norm) or the Euclidean norm of weights (L2 norm).

For coarse-grained pruning, the loss value $L_{FB}$ of each block is computed by aggregating the pruning criterion over all elements within the block, which can be formulated as:
\begin{equation}
L_{FB}(W, i, j) = \sum_{x=i}^{i+m-1}\sum_{y=j}^{j+n-1} \rho(W[x, y]),
\label{eq:fullblock_loss}
\end{equation}
where $W$ is the weight matrix, $(i, j)$ represents the top-left corner of the block, $(m, n)$ denotes the block size, and $\rho(\cdot)$ is the pruning criterion.
For fine-grained pruning, the workflow prunes each block according to a predefined sparse pattern set $\mathcal{P}$.
The loss value $L_{IB}$ for each candidate pattern $P_k \in \mathcal{P}$ is calculated by aggregating the pruning criterion over the elements to be pruned according to the pattern:
\begin{equation}
L_{IB}(W, i, j, P_k) = \sum_{(x, y) \in \Omega_k} \rho(W[i+x, j+y]),
\label{eq:intrablock_loss}
\end{equation}
where $\Omega_k = \{(x, y) | P_k[x, y] = 0\}$ represents the set of indices corresponding to the pruned elements in pattern $P_k$.
The blocks and patterns with the lowest loss values are then selected for pruning.

The pruning workflow iterates over the DNN layers and generates pruning masks for each layer based on user-defined block configurations and pruning strategies.
The generated masks are then applied to the weight matrix to zero out the pruned elements, resulting in a sparse model that aligns with the structural constraints of the target CIM architecture.
By integrating the \textit{FlexBlock} abstraction with a configurable pruning workflow, CIMinus enables users to explore a wide range of sparsity patterns and pruning strategies for CIM-based DNN acceleration.

%% file: texfiles/5-Methodology.tex
\section{Modeling Methodology}\label{sec:methodology}

Accurate modeling of latency and energy consumption is crucial for evaluating the performance of CIM architectures. 
In this section, we present a comprehensive methodology for estimating these key metrics, and discuss the overhead associated with sparsity support in CIM systems.

\subsection{Latency and Energy Modeling}
As outlined in Sec.~\ref{sec:framework:overview}, the latency of a CIM system is determined by the critical path among its hardware components, while the energy consumption can be estimated by the sum of per-access and per-cycle energy of each hardware unit.
Here we describe the modeling methodology for these metrics in detail.

\textbf{Latency.}
The estimation of latency entails calculating the number of cycles required by a unit for a given workload, taking into account the overlap in execution times with other units.
To determine overall latency, the calculation must account for the longest execution period among units operating in parallel, and the durations of operations that do not overlap.
Specifically, the execution of DNN workloads on CIM architectures involves loading the weight and feature data, performing computations for each operation, and writing back the operation results.
Contemporary CIM designs often employ a pipelined approach to these stages for maximal parallelism.
Therefore, the overall latency $L^{\text{total}}$ can be estimated by:
\begin{equation}
    L^{\text{total}} = L_{1}^{\text{load}} + \sum_{i=2}^{n} P_i(L_{i}^{\text{load}}, L_{i-1}^{\text{comp}}, L_{i-1}^{\text{wb}}) + L_{n}^{\text{comp}} + L_{n}^{\text{wb}},
\end{equation}
where $n$ denotes the total steps required in the pipeline for the DNN workload, $L_{i}^{\text{load}}$, $L_{i}^{\text{comp}}$, $L_{i}^{\text{wb}}$ represent the latencies for data loading, computation, and write-back during the $i$-th step. 
The function $P_i(\cdot)$ calculates the latency of each intermediate step within the pipeline, accounting for potential overlaps among these stages enabled by the architectural design and buffer capacities.
Therefore, the result of $P_i(\cdot)$ might be $L^{\text{load}}$ or $L^{\text{comp}}$ if loading or computation respectively constitutes the bottleneck, or $L^{\text{comp}} + L^{\text{wb}}$ if the buffer constraints prevent the parallel execution of these stages.

\begin{figure*}[!t]
    \centering
    \subfigure[Results correlation.]{
        \includegraphics[width=0.13\textwidth]{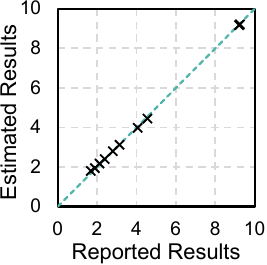}
        \label{fig:validation:correlation}
    }
    \subfigure[Speedups and energy savings. R and E denote reported and estimated results, respectively.]{
        \includegraphics[width=0.32\textwidth]{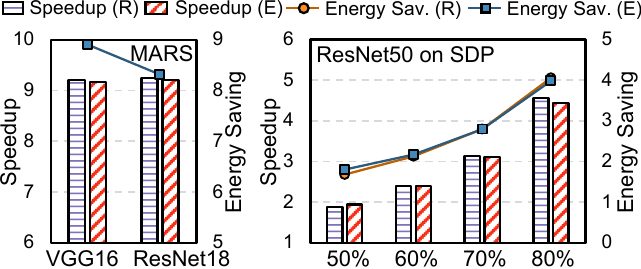}
        \label{fig:validation:results}
    }
    \subfigure[Power breakdown validation for SDP.]{
    \includegraphics[width=0.155\textwidth]{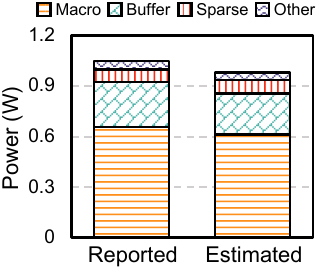}
    \label{fig:validation:breakdown}
    }
    \subfigure[Accuracy comparison against reported results for sparse DNN models.]{
        \includegraphics[width=0.315\textwidth]{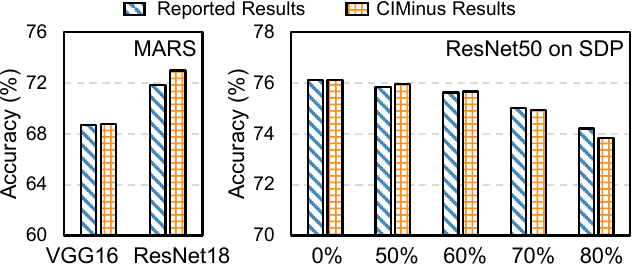}
        \label{fig:validation:accuracy}
    }
    \caption{Validation results against MARS and SDP.}
    \vspace{-10pt}
    \label{fig:validation}
\end{figure*}

\textbf{Energy Consumption.}
The overall energy consumption $E^{\text{total}}$ is the sum of computation energy of each compute unit $E^{\text{comp}}$, the read/write energy of each memory structure $E^{\text{mem}}$, and the static energy of all units $E^{\text{stat}}$:
\begin{equation}
    E^{\text{total}} = \sum_i E^{\text{comp}[i]} + \sum_j E^{\text{mem}[j]} + \sum_k E^{\text{stat}[k]}.
\end{equation}

The computation energy of each compute unit is the product of the energy per access $E^{\text{acc}}$ and the number of access $N_{\text{acc}}$:
\begin{equation}
    E^{\text{comp}[i]} = E^{\text{acc}[i]} \times N^{\text{comp}[i]}_{\text{acc}},
\end{equation}
while the read/write energy of each memory unit is the product of the energy per read ($E^{\text{read}}$) or write ($E^{\text{write}}$) with the total number of memory reads ($N_{\text{read}}$) and writes ($N_{\text{write}}$):
\begin{equation}
    E^{\text{mem}[j]} = E^{\text{read}[j]} \times N^{\text{mem}[j]}_{\text{read}} + E^{\text{write}[j]} \times N^{\text{mem}[j]}_{\text{write}}.
\end{equation}
The static energy of each unit can be derived from the product of the overall latency $L^{\text{total}}$ and the static power $P^{\text{stat}}$:
\begin{equation}
    E^{\text{stat}[k]} = P^{\text{stat}[k]} \times L^{\text{total}}.
\end{equation}

CIMinus relies on users to input the relevant parameters from $E^{\text{acc}}$, $E^{\text{read}}$, $E^{\text{write}}$, and $P^{\text{stat}}$ for each type of unit, which are generally derived from synthesis flows and memory simulation tools.
In contrast, the access counts and overall system latency are determined internally by CIMinus through cycle-level simulation.
By automatically inferring the number of units required according to the CIM array size, unit size, and the organization parameter, CIMinus simplifies the hardware description process for users while ensuring that the overall access count is accurately reflected during simulation.
Alternatively, CIMinus also provides a preset of energy parameters obtained from PTPX and PCACTI~\cite{shafaei2014fincacti}, which serves as examples for basic CIM architecture configurations.
These parameters can be useful for users who wish to conduct preliminary software-level explorations on generic CIM architectures without the need for detailed hardware specifications.

\subsection{Overhead for Sparsity Support}

Managing compressed weight matrices with \textit{FlexBlock} sparsity in CIM architectures requires storing indices for weight blocks in the finest-grained sparsity pattern, whether \textit{FullBlock} or \textit{IntraBlock}, while indices for non-zero elements are only needed for blocks with \textit{IntraBlock} sparsity.
Therefore, the additional index storage overhead $S^{\text{idx}}$ for weight matrix $W$ can be formulated as:
\begin{equation}
    S_W^{\text{idx}} = N^{W}_{\text{nz}} \times S^{\text{idx}}_B + \sum_{i=1}^{N_{\text{nz}}} N^{B_i}_{\text{nz}} \times S_{\text{elem}}^{\text{idx}} \ ,
\end{equation}
where $N^{W}_{\text{nz}}$ denotes the number of non-zero blocks in $W$, $S^{\text{idx}}_B$ represents index storage requirement for a single non-zero block, $N^{B_i}_{\text{nz}}$ is the number of non-zero elements within the $i$-th block, and $S_{\text{elem}}^{\text{idx}}$ is the index storage requirement for each non-zero element within a block.

In CIMinus, the operational costs associated with sparsity support include memory accesses for index retrieval, multiplexer operations in indexing units for input selection, additional accumulation for misaligned partial results, and the zero detection and skipping for input sparsity in pre-processing units.
Similar to the modeling of other hardware units, the framework automatically determines the required hardware resources, such as the number of multiplexers or accumulators, according to the array dimensions, organization and sparsity specifications. 
During simulation, CIMinus tracks these sparsity-related operations through the aforementioned energy and latency modeling and incorporate them into the final report, enabling designers to assess the overhead and make more informed design decisions.

%% file: texfiles/6-Evaluation.tex
\section{Framework Evaluation}\label{sec:evaluation}
In this section, we first validate CIMinus against recent CIM designs, and then perform runtime analysis to demonstrate its capability in supporting rapid design iteration.

\subsection{CIMinus Validation}\label{sec:evaluation:validation}

We validate CIMinus against results reported in recent digital SRAM-based CIM designs, namely MARS~\cite{sie2021mars} and SDP~\cite{tu2022sdp}. 
As detailed in Tab.~\ref{tab:cim-designs}, these accelerators feature varying macro designs and organizational layout, and utilize distinct sparsity patterns in DNN weights.
During simulation, we use the configurations reported in these works, whenever possible.
The power model of the CIM array is adopted from~\cite{yan20221}, and the per-access / per-cycle energy of buffers is obtained through PCACTI~\cite{shafaei2014fincacti}.
The remaining digital modules are implemented with Verilog HDL and synthesized by Design Compiler, while the power consumption is obtained by PTPX.
We evaluate the effectiveness of CIMinus in estimating key metrics for sparse DNN workloads on CIM architectures, including inference speedup, energy saving, and model accuracy.
To maintain consistency with the reported results, our validation process utilizes DNN models such as VGG16, ResNet18, and ResNet50, employing the same datasets as those used in the respective designs (CIFAR100 for MARS and ImageNet for SDP).

\begin{table}[t]
\centering
\caption{Summary of CIM designs for validation.}
\label{tab:cim-designs}
\begin{tabular}{c|cc}
\hline
\multirow{2}{*}{\textbf{Parameters}} & \multicolumn{2}{c}{\textbf{CIM Design}} \\ \cline{2-3}
 & \multicolumn{1}{c|}{MARS~\cite{sie2021mars}} & SDP~\cite{tu2022sdp} \\
\hline
Macro Size & \multicolumn{1}{c|}{$1024 \times 64$} & $32\times 64$ \\
Sub-array Size & \multicolumn{1}{c|}{$64 \times 64$} & $1\times 64$ \\
Macro Org. & \multicolumn{1}{c|}{8 macros ($2 \times 4$)} & 512 macros ($16 \times 32$) \\
Global Buf. & \multicolumn{1}{c|}{128KB (Ping-pong)} & 256KB (In), 128KB (Out) \\
Sparsity & \multicolumn{1}{c|}{\textit{Full} (1, 16)} & \textit{Intra} (2, 1) + \textit{Full} (2, 8) \\
Eval. Scope & \multicolumn{1}{c|}{Only Conv layers} & Entire NN \\
\hline
\end{tabular}
\end{table}

As illustrated in Fig.~\ref{fig:validation}, CIMinus accurately estimates both system latency and component-level energy, closely aligning with the results reported in the literature.
The correlation plot in Fig.~\ref{fig:validation:correlation} shows a strong agreement between the reported and estimated values for both speedups and energy savings, with all data points falling within a 5.27\% error margin.
This indicates that CIMinus can reliably model the performance of CIM accelerators across different DNN models and sparsity patterns.
Fig.~\ref{fig:validation:results} and \ref{fig:validation:breakdown} provide a more detailed comparison between the reported and estimated values, showing a close consistency across speedups, energy savings, and detailed power breakdowns. 
These results highlight the adaptability and precision of CIMinus in modeling various aspects of digital CIM architectures, from system-level metrics to fine-grained component-level details.
Moreover, as shown in Fig.~\ref{fig:validation:accuracy}, the DNN model accuracies obtained from the pruning workflow in CIMinus are also consistent with the reported values, demonstrating its effectiveness in acquiring sparse models based on the \textit{FlexBlock} abstraction.

While CIMinus provides accurate overall estimations of speedups, energy savings, and system-level power consumption, minor discrepancies between the reported and estimated results can be attributed to two main factors.
First, some specific design parameters, such as buffer bandwidth, are not always disclosed in the existing literature, leading to potential differences in the modeling assumptions.
Second, as a system-level tool focused on rapid design exploration, CIMinus operates at a higher abstraction level than circuit-level simulations, potentially missing optimizations that differentiate designs at the implementation level.
Despite these challenges, CIMinus still maintains a high level of accuracy, demonstrating its reliability and effectiveness in modeling sparse workloads on CIM architectures.
Future work could extend CIMinus with more detailed interface and modeling capabilities, allowing users to configure circuit-level optimizations and architectural details for more accurate performance estimation when needed.

\subsection{Runtime Analysis}\label{sec:evaluation:runtime}

\begin{figure}[t]
    \centering
    \includegraphics[width=\linewidth]{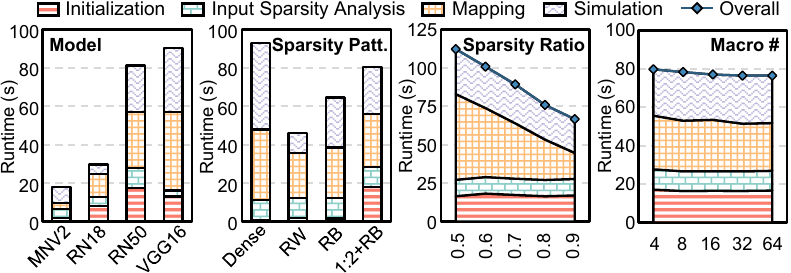}
    \vspace{-18pt}
    \caption{Framework runtime and scalability analysis across different models, sparsity patterns, sparsity ratios and macro counts. Here RW and RB denote row-wise and row-block sparsity pattern, respectively.}
    \label{fig:runtime}
\end{figure}

We evaluate the runtime efficiency of CIMinus to demonstrate that designers can rapidly evaluate multiple design configurations within practical time constraints.
We analyze runtime using a 4-macro configuration with 80\% sparsity and input sparsity enabled, across models ranging from 3.4M (MobileNetV2) to 138M (VGG16) parameters with hybrid 1:2 and row-block sparsity, and across various sparsity patterns on ResNet50.
As shown in Fig.~\ref{fig:runtime}, the runtime remains consistently under 100 seconds over these configurations, which is generally dominated by operator mapping and cycle-level simulation.
The initialization overhead also increases noticeably for fine-grained \textit{IntraBlock} patterns due to the complexity of element-level index extraction.
Similarly, the input sparsity analysis overhead increases with deeper networks, as the input activation extraction is required for every layer.

We further examine framework scalability by varying sparsity ratios from 0.5 to 0.9 and macro counts from 4 to 64, using ResNet50 with the same hybrid sparsity pattern.
The results in Fig.~\ref{fig:runtime} show that the runtime scales primarily with workload complexity rather than hardware scale, as mapping and simulation must process each operation regardless of available parallel resources.
This scaling behavior aligns well with sparse workload evaluation, as sparsity reduces the number of operations to process, enabling efficient exploration across diverse hardware configurations.

%% file: texfiles/7-Exploration.tex
\section{Exploration with CIMinus}\label{sec:exploration}

In this section, we demonstrate the practical utility of CIMinus through two illustrative use cases in exploring and evaluating diverse design choices.
Specifically, we use CIMinus to investigate the impact of sparsity exploitation on DNN model accuracy and system efficiency, and discuss the performance implications of various mapping strategies.

\subsection{Experimental Setup}\label{sec:exploration:setup}

\begin{table}
\centering
\caption{Summary of sparsity patterns and their corresponding FlexBlock representations.}
\label{tab:sparsity-patterns}
\begin{tabular}{c|l}
\hline
\textbf{Sparsity Pattern} & \textbf{\textit{FlexBlock} Representation} \\ 
\hline
Row-wise & \textit{FullBlock} (1, $N$) \\
Row-block & \textit{FullBlock} (1, 16) \\
Column (Filter)-wise & \textit{FullBlock} ($M$, 1) \\
Channel-wise & \textit{FullBlock} ($C_{in}$, 1) \\
Column-block & \textit{FullBlock} (16, 1) \\
1:2 + Row-block & \textit{IntraBlock} (2, 1) + \textit{FullBlock} (2, 16) \\
1:2 + Row-wise & \textit{IntraBlock} (2, 1) + \textit{FullBlock} (2, $N$) \\
1:4 + Row-block & \textit{IntraBlock} (4, 1) + \textit{FullBlock} (4, 16) \\ 
\hline
\end{tabular}
\end{table}

We use CIMinus to conduct two in-depth exploration studies examining sparsity exploitation and mapping strategies.
Both studies employ a common CIM architecture configuration with 8-bit precision for both weights and features.
Each macro contains an array of $1024\times 32$ with sub-arrays of $32\times 32$, utilizing a weight stationary mapping where weight matrix rows are unrolled along the array row dimension.
For sparsity exploitation analysis, we configure a 4-macro architecture where all macros share broadcasted inputs from a single input buffer.
We evaluate the trade-off between model accuracy and efficiency by comparing sparse workloads with weight and input sparsity against a dense baseline, which executes dense workloads using the same architecture configuration without specialized hardware support for sparsity.
For mapping strategy exploration, we extend the architecture to 16 macros while maintaining the same per-macro specifications.
This enables the exploration of diverse macro organizations, such as $8 \times 2$, $4 \times 4$, and $2 \times 8$, where the organization dimensions offer flexibility for either spatial mapping of weight matrix rows or weight duplication strategies.
The mapping exploration examines how different mapping approaches impact overall latency, energy consumption, and array utilization.

To investigate the impact of weight sparsity patterns on model accuracy and system efficiency, we use CIMinus to evaluate representative sparsity patterns across ResNet50, VGG16, and MobileNetV2, as summarized in Tab.~\ref{tab:sparsity-patterns}.
These patterns span various granularities, including row-wise, column-wise, and more fine-grained \textit{FullBlock} sparsity patterns, as well as hybrid patterns that combine \textit{IntraBlock} and \textit{FullBlock} sparsity.
Here, $M$, $N$, and $C_{in}$ denote the number of rows and columns in the weight matrices and the size of the input channel, respectively.
The sparsity ratio of these patterns ranges from 0.5 to 0.9, allowing us to observe the efficiency-accuracy trade-off at different sparsity levels.
For hybrid patterns, the \textit{IntraBlock} sparsity ratio is fixed such that only one element within each block remains non-zero, while the \textit{FullBlock} sparsity ratio is adjusted accordingly to maintain the overall sparsity ratio.
Additionally, we evaluate the benefits of input sparsity across these three models, examining both its impact on dense models and its interaction with different weight sparsity patterns and ratios on ResNet50.

\subsection{Sparsity Exploitation Analysis}\label{sec:exploration:sparsity}

\begin{figure}[t]
    \centering
    \subfigure[Speedups.]{
        \hspace{-5pt}
        \includegraphics[width=0.97\linewidth]{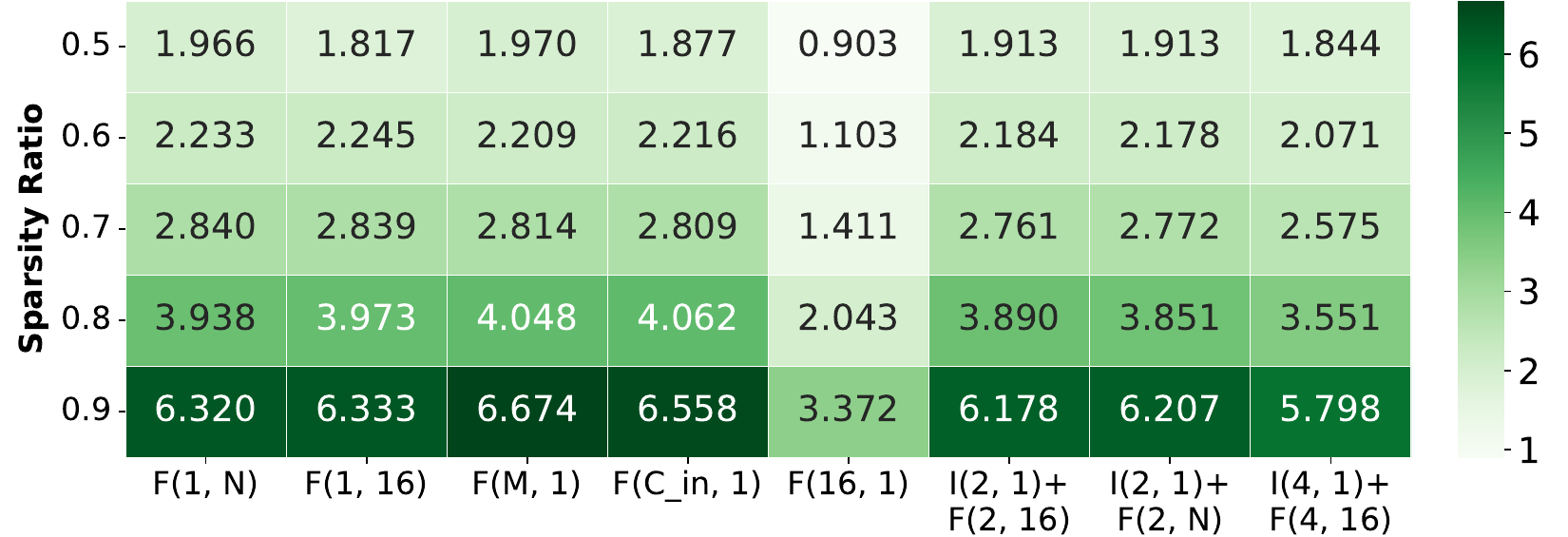}
        \label{fig:explore_sp:speedup}
    }
    \subfigure[Energy savings.]{
        \hspace{-5pt}
        \includegraphics[width=0.97\linewidth]{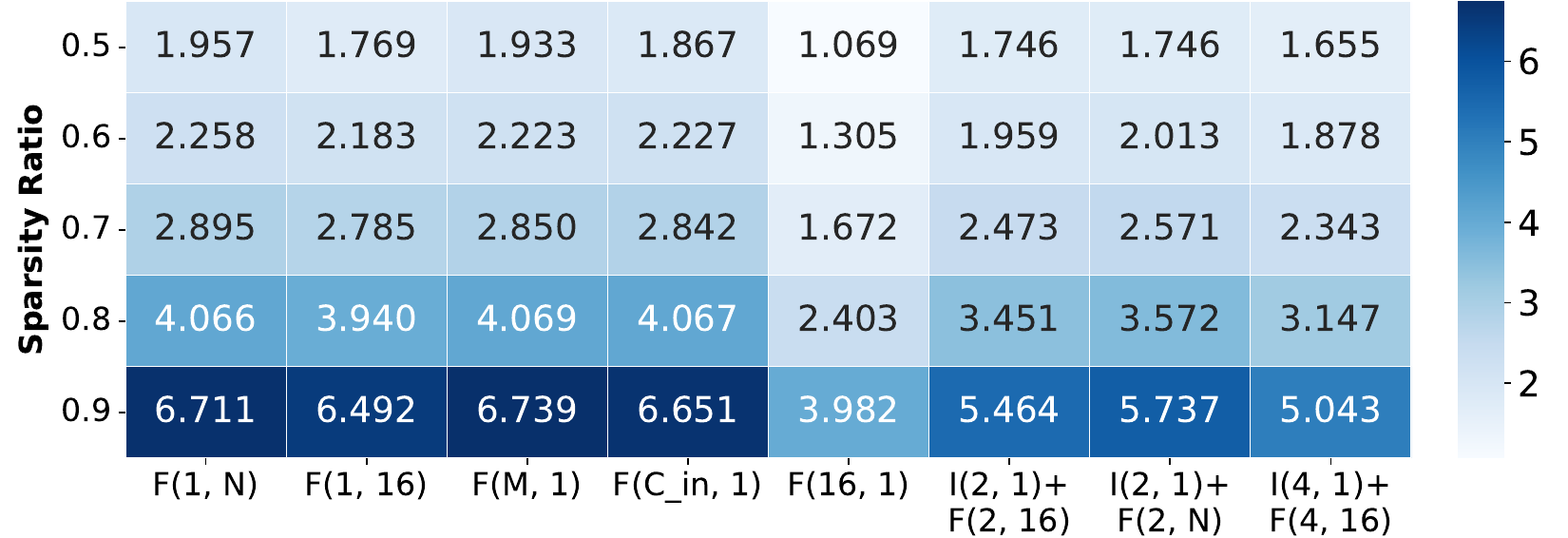}
        \label{fig:explore_sp:energy}
    }
    \subfigure[Model accuracies.]{
        \hspace{-5pt}
        \includegraphics[width=0.97\linewidth]{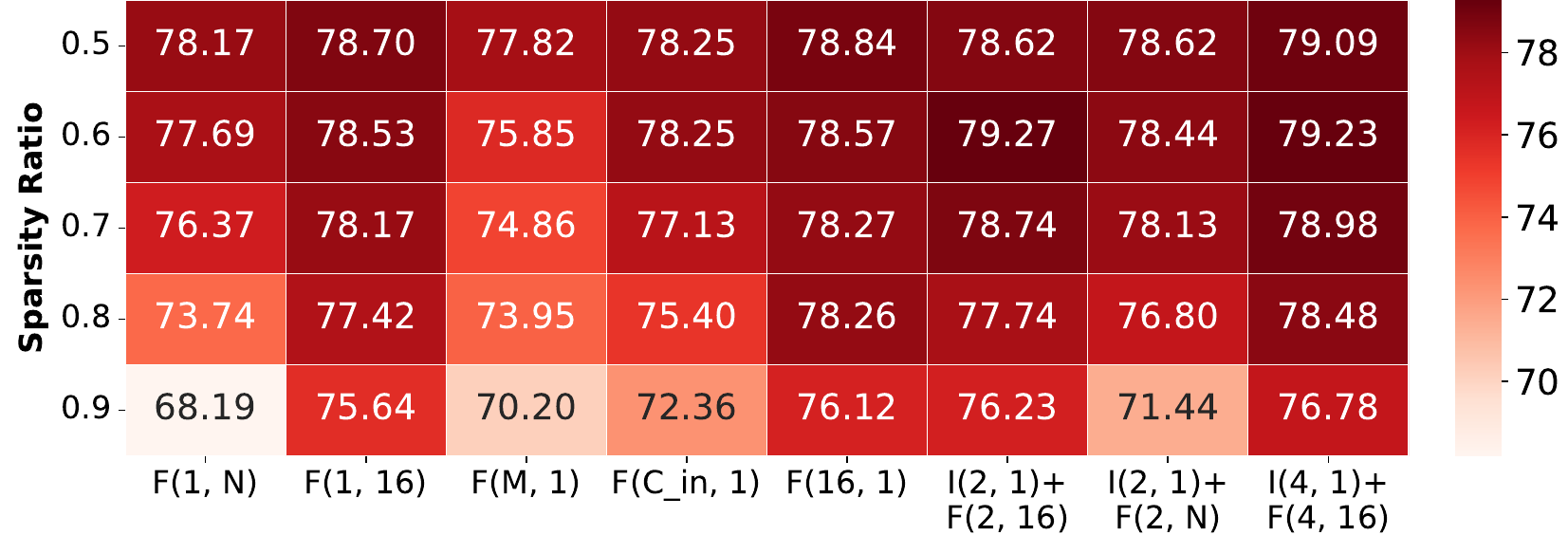}
        \label{fig:explore_sp:accuracy}
    }
    \vspace{-10pt}
    \caption{Speedup, energy saving and model accuracy comparison on ResNet50 with CIFAR-100. Here, F is short for \textit{FullBlock} and I is short for \textit{IntraBlock}.}
    \label{fig:explore_sp}
\end{figure}

Fig.~\ref{fig:explore_sp} presents the speedup, energy saving, and model accuracy results compared against dense baseline implementation for the ResNet50 model with the CIFAR-100 dataset.
As the results indicate, coarser-grained sparsity patterns, such as \textit{FullBlock} patterns that apply pruning by entire dimensions, tend to bring higher efficiency improvement at the cost of considerable accuracy degradation.
In contrast, patterns with smaller block sizes typically achieve better accuracy due to their fine-grained and flexible nature, which is especially evident for hybrid patterns. 
However, these fine-grained patterns are less likely to fully align with the CIM array structure, leading to lower efficiency gains. 
Moreover, the routing logic required for supporting \textit{IntraBlock} sparsity introduces additional energy consumption overhead, which partially negates the energy benefits from sparsity exploitation.
This overhead also becomes more noticeable with increasing block sizes.

\begin{figure}[t]
    \centering
    \subfigure[Evaluation across block sizes.]{
        \includegraphics[width=\linewidth]{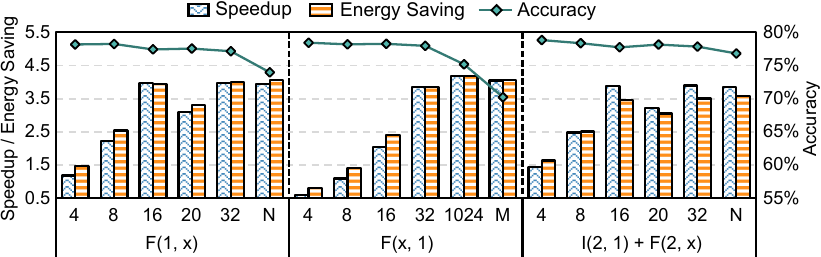}
        \label{fig:explore_sp_detail:blocksize}
    }
    \subfigure[Evaluation across NN models.]{
        \includegraphics[width=\linewidth]{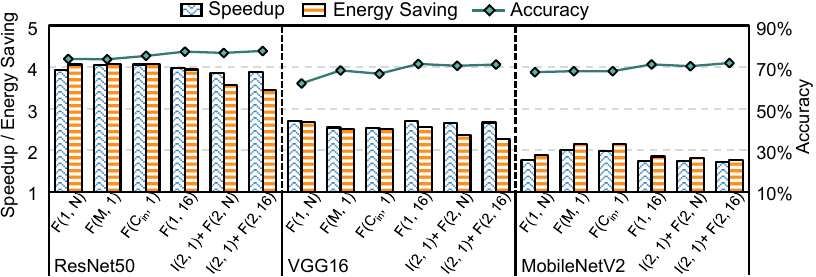}
        \label{fig:explore_sp_detail:network}
    }
    \vspace{-10pt}
    \caption{Evaluation across block sizes and neural network architectures with sparsity ratio maintained at 80\%.}
    \vspace{-8pt}
    \label{fig:explore_sp_detail}
\end{figure}

To conduct a more detailed analysis, we examine the impact of block dimensions and neural network architectures with the sparsity ratio maintained at 80\%.
Fig.~\ref{fig:explore_sp_detail:blocksize} illustrates the influence of block sizes on accuracy and system efficiency.
Our analysis focuses on three representative sparsity patterns: row-block, column-block, and hybrid sparsity.
For a comprehensive comparison, the block sizes are selected to either align with or differ from the dimensions of optimal parallelism: 16 for broadcasting input across the row dimension, and 32 for accumulation along the column dimension.
Although patterns with larger block sizes generally achieve higher accuracy and lower efficiency, when block sizes are not integral multiples of these optimal dimensions, misalignment can lead to fragmentation during weight mapping.
These results also corroborate the constraints of CIM architectures discussed in Sec.~\ref{sec:sparsity:limitation}.
Fig.~\ref{fig:explore_sp_detail:network} depicts the results across ResNet50, VGG16, and MobileNetV2.
Despite consistent overall trends across these models, the latter two show less pronounced improvements compared to their dense baselines. 
This is due to the significant accuracy drop associated with pruning the FC layers in VGG16 and the depthwise Conv layers in MobileNetV2. 
As a result, pruning was restricted to standard Conv layers, leading to reduced efficiency gains.

\begin{figure}[t]
    \centering
    \includegraphics[width=\linewidth]{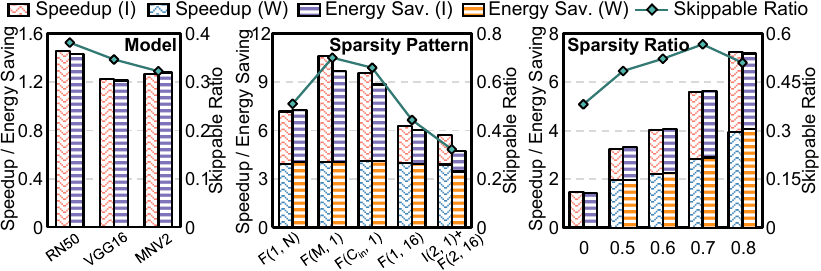}
    \vspace{-18pt}
    \caption{Evaluation of input sparsity exploitation across different models, weight sparsity patterns and weight sparsity ratios. I and W denote with and without input sparsity support enabled, respectively.}
    \label{fig:input_sparsity}
\end{figure}

Beyond weight sparsity, we also use CIMinus to evaluate the benefits of input sparsity exploitation by skipping ineffectual bit-serial computations.
We analyze both the ratio of skippable computations determined through workload profiling, and how well they translate to speedups and energy savings with input sparsity support enabled. 
As shown in Fig.~\ref{fig:input_sparsity}, input sparsity provides consistent improvements across all evaluated dense workloads, achieving speedups and energy savings ranging from 1.2$\times$ to 1.4$\times$.
To examine the combined benefits of input and weight sparsity, we evaluate their interaction across different weight sparsity patterns at 80\% sparsity, and across various sparsity ratios using the row-wise pattern.
The results show that the skippable ratio varies significantly with weight sparsity configurations.
Coarse-grained patterns such as column-wise or channel-wise patterns achieve high skippable ratios, while \textit{IntraBlock} patterns are less likely to skip computations due to multiple inputs being processed simultaneously per CIM row.
Moreover, the speedups and energy savings brought by input sparsity also increase with sparser models, potentially due to shifts in activation distributions that lead to increased zero activations in the bit-serial representation.
These results demonstrate that input sparsity further amplifies the efficiency advantages of coarse-grained patterns, reinforcing the trade-off observed with weight sparsity alone.

\begin{finding}{Finding 1:}
In CIM systems, the high efficiency gains from coarser-grained sparsity patterns come with a significant accuracy trade-off. 
Finer-grained patterns, especially when aligned with the hardware architecture, can achieve a more optimal balance between efficiency and accuracy.
\end{finding}
\vspace{-15pt}

\subsection{Mapping Strategies Exploration}

\begin{figure}[t]
    \centering
    \includegraphics[width=\linewidth]{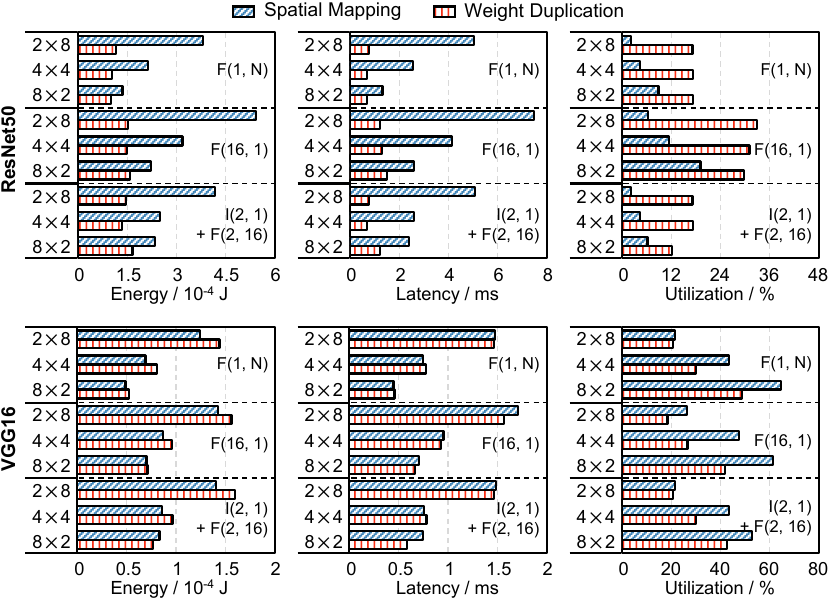}
    \caption{Energy, latency, and array utilization of different mapping strategies for ResNet50 and VGG16 across various macro organizations.}
    \vspace{-8pt}
    \label{fig:explore_mp_dup}
\end{figure}

\begin{figure}[t]
    \centering
    \subfigure[Energy breakdown.]{
        \includegraphics[width=0.255\textwidth]{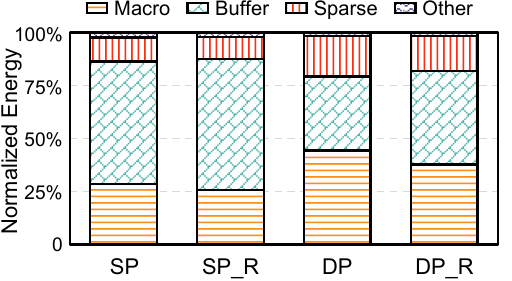}
        \label{fig:explore_mp_slice:breakdown}
    }
    \subfigure[Performance comparison.]{
        \includegraphics[width=0.195\textwidth]{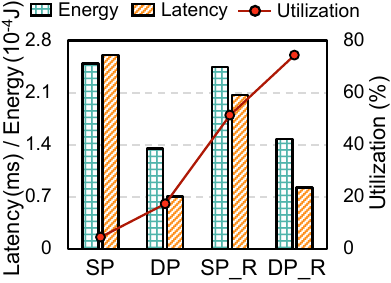}
        \label{fig:explore_mp_slice:slicing}
    }
    \caption{Normalized energy breakdown, along with energy, latency, and utilization comparison with (denoted by R) and without weight data rearrangement. Here SP and DP denote spatial mapping and weight duplication.}
    \label{fig:explore_mp_slice}
\end{figure}

Fig.~\ref{fig:explore_mp_dup} illustrates the performance of spatial mapping and weight duplication for ResNet50 and VGG16 across macro organizations.
In general, highly compressed sparse weight matrices in Conv layers often fail to fully occupy CIM arrays, leading to under-utilization. 
This inefficiency is particularly pronounced when the designated dimension is 8.
For models dominated by Conv layers, such as ResNet50, weight duplication can significantly improve array utilization, achieving up to a $7.7\times$ increase.
Notably, the evenly distributed dimensions in the $4 \times 4$ organization appear to facilitate more balanced parallelism for both weights and features, resulting in optimal energy efficiency and latency when combined with duplication.
In contrast, models with a large proportion of FC layer parameters, such as VGG16, experience decreased performance and utilization with weight duplication.
This may be attributed to the nature of FC layers, which typically exhibit less weight data reuse, thus limiting the applicability of duplication strategies.

To evaluate the effect of weight data rearrangement, we compare the energy, latency, and array utilization with and without the equalization of ragged compressed sparse weight matrices.
This evaluation employs a hybrid pattern combining \textit{IntraBlock} (2, 1) and \textit{FullBlock} (2, 16), while utilizing a $4 \times 4$ macro organization.
As illustrated in Fig.~\ref{fig:explore_mp_slice}, weight rearrangement substantially improves array utilization through a more balanced workload distribution. 
However, this does not guarantee an overall performance improvement. 
The energy efficiency gained in CIM macros is counterbalanced by increased buffer access overhead, revealing that structural constraints can limit the benefits of higher utilization.

\begin{finding}{Finding 2:}
Duplicating compressed weights and balancing the workload among macros can significantly enhance parallelism and utilization. 
However, while higher array utilization generally improves efficiency, it does not always guarantee better efficiency due to structural constraints.
\end{finding}
\vspace{-10pt}

%% file: texfiles/8-RelatedWorks.tex
\section{Related Works}\label{sec:related}

In recent years, SRAM-based CIM architectures have explored various sparsity patterns to reduce computation and storage requirements of DNN workloads~\cite{wang2022spcim, yue202014, yue202115, sie2021mars, tu2022sdp, kim2021z}.
Early analog CIM designs prune weights along the kernel or channel direction and adopt block-wise zero skipping to improve energy efficiency~\cite{yue202014, yue202115}.
As the field shifts toward digital CIM implementations, their full-array activation imposes stricter structural constraints on sparsity patterns.
MARS~\cite{sie2021mars} addresses this through group-wise structured pruning and index-aware optimizations that align with the array dimensions.
SDP~\cite{tu2022sdp} employs double-broadcast hierarchical pruning that combines 1:2 sparsity with row-wise patterns, supported by dedicated routing and index storage.
While these designs demonstrate effective sparsity exploitation in CIM, the vast design space still remains largely unexplored.

To enable rapid design space exploration of CIM-based accelerators, researchers have developed several evaluation frameworks for modeling the hardware behavior and performance at various design levels.
DNN+NeuroSim~\cite{peng2019dnn+} is an end-to-end benchmarking framework that integrates circuit-level hardware models with PyTorch and TensorFlow, focusing on trade-offs brought by device non-idealities.
MNSIM 2.0~\cite{zhu2023mnsim} is a behavior-level modeling tool that features a unified array model for both analog and digital CIM, supported by a specialized model training and quantization flow.
ZigZag-IMC~\cite{sun2023analog} integrates an analytical CIM performance model into a system-level exploration framework to quantitatively benchmark and compare analog and digital CIMs with various DNN workloads.
CiMLoop~\cite{andrulis2024cimloop} performs efficient system-level modeling of CIM co-designs through the combination of a flexible hardware specification and a fast statistical energy model.
Although existing frameworks have significantly accelerated CIM design and evaluation compared to circuit-level simulations, these frameworks exclusively target dense workloads, lacking the abstractions and modeling capabilities needed for sparsity exploitation.
Without support for diverse sparsity patterns, irregular tensor compression, or the associated indexing and routing logic, these tools cannot effectively explore the growing design space of sparse CIM architectures.
CIMinus bridges this gap in existing evaluation frameworks by introducing the \textit{FlexBlock} abstraction for effectively representing structurally constrained sparsity patterns, and providing an integrated workflow from model-pruning to system-level evaluation validated against real sparse CIM designs. 

%% file: texfiles/9-Conclusion.tex
\section{Conclusion}\label{sec:conclusion}
In this paper, we present CIMinus, a framework designed for modeling sparse DNN workloads on SRAM-based CIM architectures.
CIMinus accepts high-level descriptions of sparse DNN workloads, hardware designs, and mapping methods, and provides precise estimations of latency and energy based on these inputs.
We validate CIMinus against results from recent SRAM-based CIM designs, and showcase its utility in facilitating the exploration of trade-offs in sparsity pattern and mapping strategy selections through two illustrative use-cases. 
We envision that CIMinus will serve as a valuable tool for designers, aiding in the development of efficient CIM systems for a wide range of applications.

%% file: texfiles/bio.tex
\vspace{-17mm}

\begin{IEEEbiography}[{\includegraphics[width=1in,height=1.25in,clip,keepaspectratio]{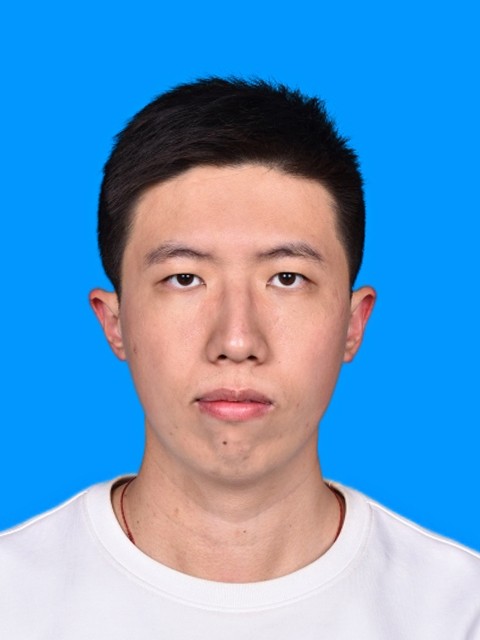}}]{Yingjie Qi}

received the B.S. degree in computer science and technology from Beihang University, Beijing, China, in 2020. He is currently pursuing the Ph.D. degree at the School of Computer Science and Engineering, Beihang University, China. His research interests include compute-in-memory architectures, deep learning compilers, and graph neural networks acceleration.

\end{IEEEbiography}

\vspace{-15mm}

\begin{IEEEbiography}[{\includegraphics[width=1in,height=1.25in,clip,keepaspectratio]{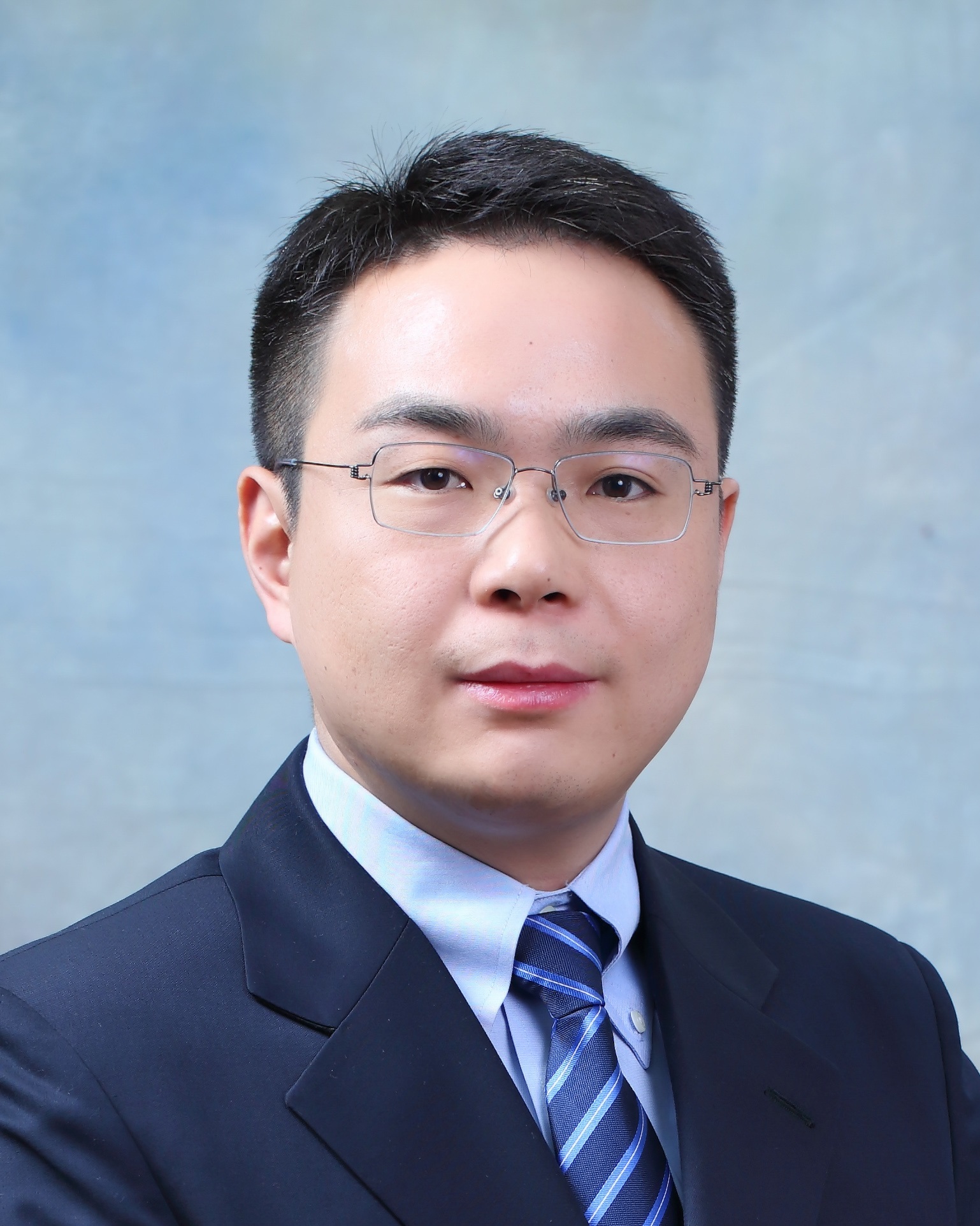}}]{Jianlei Yang}

(S'11-M'14-SM'20) received the B.S. degree in microelectronics from Xidian University, Xi'an, China, in 2009, and the Ph.D. degree in computer science and technology from Tsinghua University, Beijing, China, in 2014.

He is currently a Professor in Beihang University, Beijing, China, with the School of Computer Science and Engineering. From 2014 to 2016, he was a post-doctoral researcher with the Department of ECE, University of Pittsburgh, Pennsylvania, USA.
His current research interests include emerging computer architectures, hardware-software co-design and machine learning systems.

Dr. Yang was the recipient of the First/Second place on ACM TAU Power Grid Simulation Contest in 2011 and 2012. He was a recipient of IEEE ICCD Best Paper Award in 2013, ACM GLSVLSI Best Paper Nomination in 2015, IEEE ICESS Best Paper Award in 2017, ACM SIGKDD Best Student Paper Award in 2020.

\end{IEEEbiography}

\vspace{-20mm}

\begin{IEEEbiography}[{\includegraphics[width=1in,height=1.25in,clip,keepaspectratio]{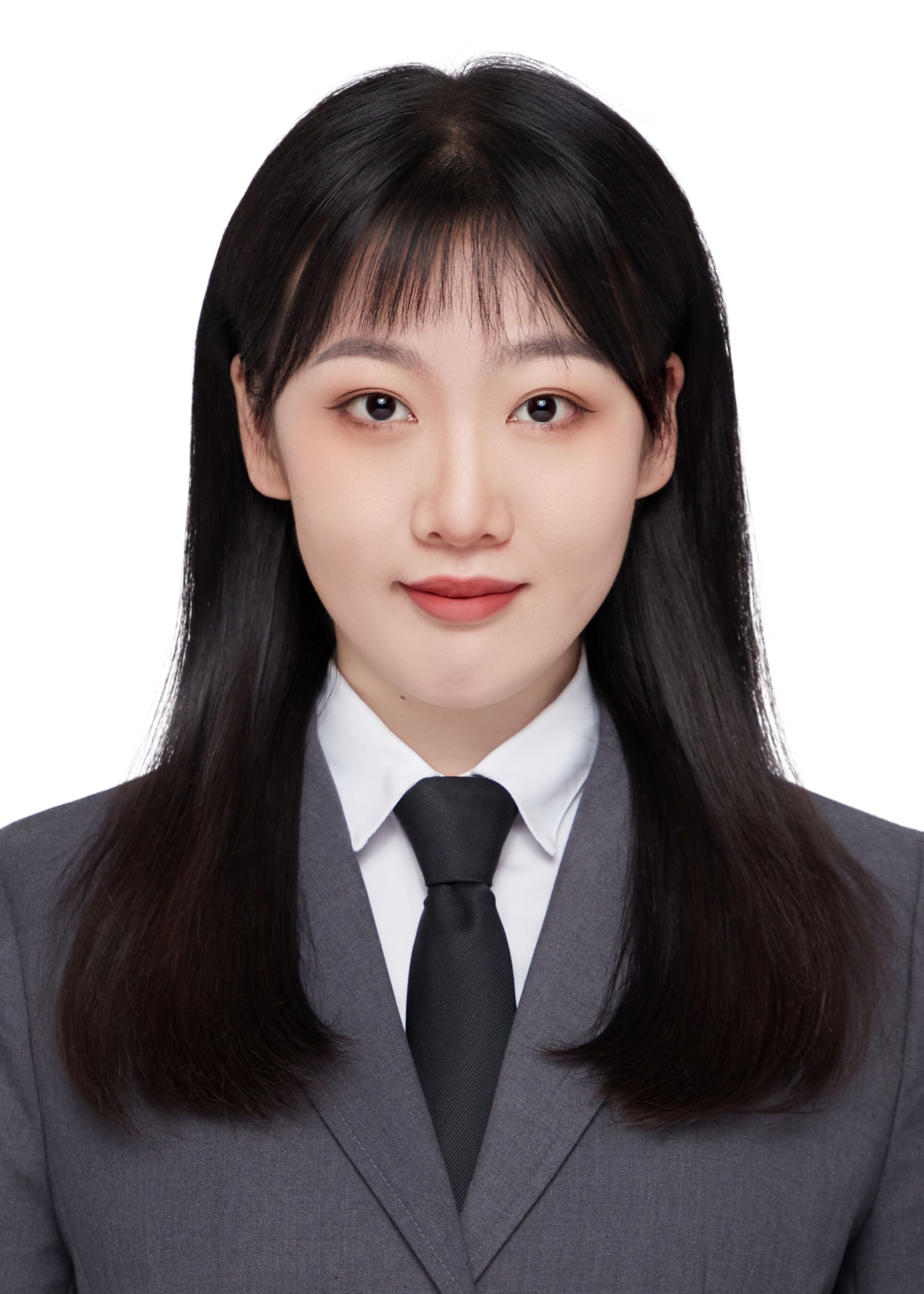}}]
{Rubing Yang}

received the B.S. degree in computer science and technology from Beihang University, Beijing, China, in 2023. She is currently pursuing the M.S. degree at the School of Computer Science and Engineering, Beihang University, China. Her research interests include model compression, processing-in-memory architectures, and large language model inference optimization.

\end{IEEEbiography}

\vspace{-15mm}

\begin{IEEEbiography}[{\includegraphics[width=1in,height=1.25in,clip,keepaspectratio]{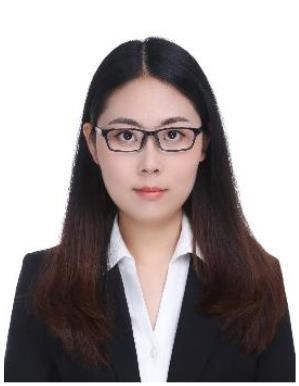}}]{Cenlin Duan}

received the B.S. degree in Electronic Science and Technology from University of Electronic Science and Technology of China, Chengdu, China, in 2015, and the M.S. degree in Software Engineering from Xidian University, Xi'an, China, in 2018. She is currently pursuing the Ph.D. degree at the School of Integrated Circuit Science and Engineering, Beihang University, Beijing, China. Her current research interests include processing-in-memory architectures and deep learning accelerators.

\end{IEEEbiography}

\vspace{-15mm}

\begin{IEEEbiography}[{\includegraphics[width=1in,height=1.25in,clip,keepaspectratio]{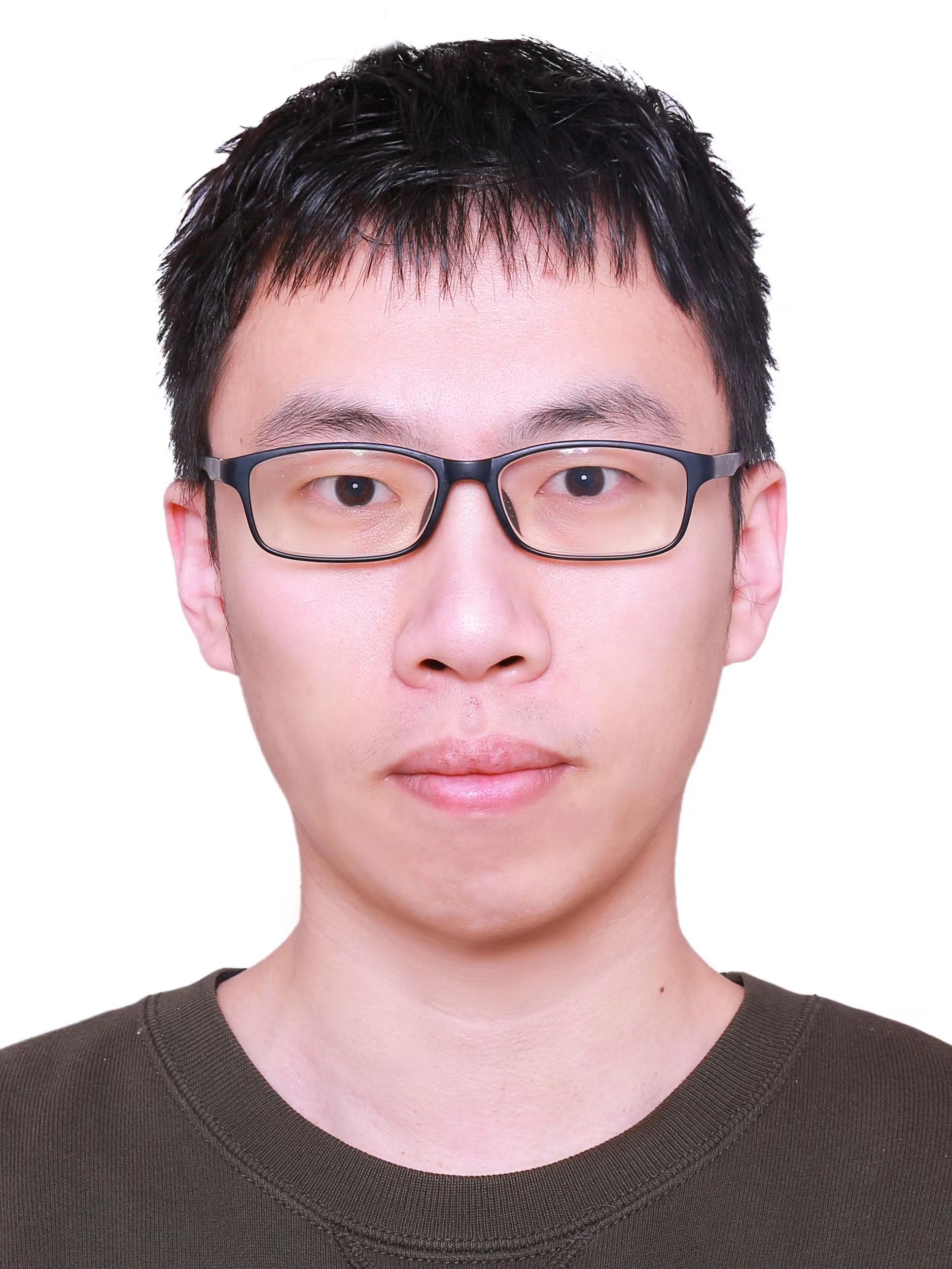}}]{Xiaolin He}

received the B.S. degree in software engineering from Beihang University, Beijing, China, in 2020. He is currently pursuing the Ph.D. degree at the School of Computer Science and Engineering, Beihang University, China. His research interests include in-memory computing architectures and compiler optimization techniques.

\end{IEEEbiography}

\vspace{-15mm}

\begin{IEEEbiography}[{\includegraphics[width=1in,height=1.25in,clip,keepaspectratio]{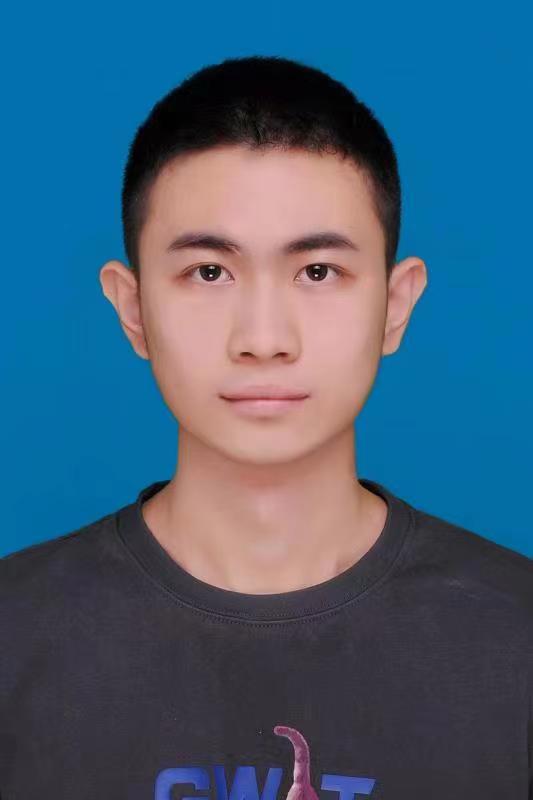}}]{Ziyan He}

received the B.S. degree in telecommunication engineering in 2022 and the M.S. degree in 2025, both from Xidian University, Xi’an, China. His current research interests include processing-in-memory architecture and domain-specified accelerators.

\end{IEEEbiography}

\vspace{-15mm}

\begin{IEEEbiography}[{\includegraphics[width=1in,height=1.25in,clip,keepaspectratio]{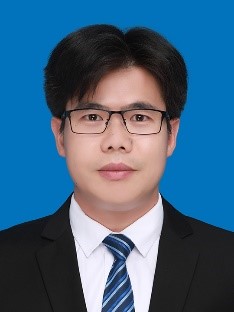}}]{Weitao Pan}

received the B.S. degree from School of Technical Physics of Xidian University in 2004. His Ph.D. degree was received from School of Microelectronics of Xidian University in 2010. Now he is an associate professor in State Key Laboratory of Integrated Service Networks of Xidian University. His current research interests include VLSI design methods and post-silicon verification.

\end{IEEEbiography}

\vspace{-15mm}

\begin{IEEEbiography}[{\includegraphics[width=1in,height=1.25in,clip,keepaspectratio]{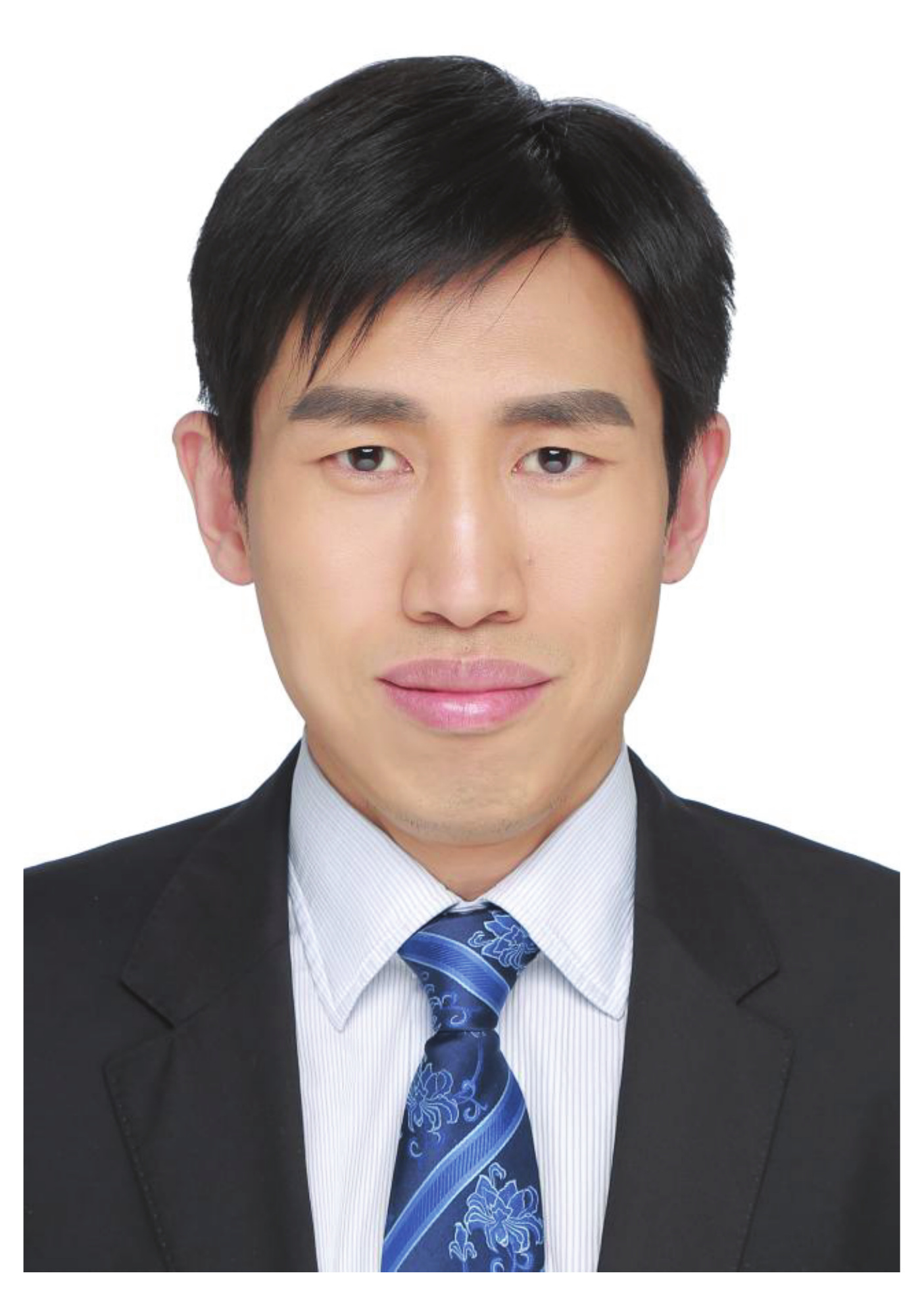}}]{Weisheng Zhao}

(Fellow, IEEE) received the Ph.D. degree in physics from the University of Paris Sud, Paris, France, in 2007.

He is currently a Professor with the School of Integrated Circuit Science and Engineering, Beihang University, Beijing, China. In 2009, he joined the French National Research Center, Paris, as a Tenured Research Scientist. Since 2014, he has been a Distinguished Professor with Beihang University. He has published more than 300 scientific articles in leading journals and conferences, such as \textit{Nature
Electronics}, \textit{Nature Communications}, \textit{Advanced Materials}, IEEE Transactions, ISCA, and DAC. His current research interests include the hybrid integration of nanodevices with CMOS circuits and new nonvolatile memory (40-nm technology node and below), like MRAM circuit and architecture design.

Prof. Zhao was the Editor-in-Chief for the {\sc{IEEE Transactions on Circuits and System I: Regular Paper}} from 2020 to 2023.

\end{IEEEbiography}